\documentclass[twocolumn,prx,longbibliography,aps,notitlepage,showpacs,amsmath,amssymb,superscriptaddress]{revtex4-1}

\usepackage{float}
\usepackage{times}
\usepackage{bm}
\usepackage{textcomp}
\usepackage{graphicx}
\usepackage{bm}
\allowdisplaybreaks 
\usepackage{braket}
\usepackage{array}
\usepackage[breaklinks]{hyperref}
\usepackage{subdepth}
\usepackage{mathtools}
\hypersetup{colorlinks=true, linkcolor=blue, citecolor=blue, filecolor=blue, urlcolor=blue}
\newcommand{\RN}[1]{\textup{\uppercase\expandafter{\romannumeral#1}}}
\begin{document}

\title{Catastrophe theory classification of Fermi surface topological transitions in two dimensions}

\author{Anirudh Chandrasekaran}
\affiliation{Department of Physics, Boston University, Boston, MA, 02215, USA.}

\author{Alex Shtyk}
\affiliation{Quantlab Group, Cambridge, Massachusetts, USA.}

\author{Joseph J. Betouras}
\affiliation{Department of Physics, Loughborough University, Loughborough LE11 3TU, UK.}

\author{Claudio Chamon}
\affiliation{Department of Physics, Boston University, Boston, MA, 02215, USA.}

\date{\rm\today}

\begin{abstract}

We classify all possible singularities in the electronic dispersion of two-dimensional systems that occur when the Fermi surface changes topology, using catastrophe theory. For systems with up to seven control parameters (i.e., pressure, strain, bias voltage, etc), the theory guarantees that the singularity belongs to to one of seventeen standard types. We show that at each of these singularities the density of states diverges as a power law, with a universal exponent characteristic of the particular catastrophe, and we provide its universal ratio of amplitudes of the prefactors of energies above and below the singularity. We further show that crystal symmetry restricts which types of catastrophes can occur at the points of high symmetry in the Brillouin zone. For each of the seventeen wallpaper groups in two-dimensions, we list which catastrophes are possible at each high symmetry point. 
\end{abstract}

\maketitle


\section{Introduction}
Valence electrons in crystalline solids are described by Bloch states with a dispersion relation $\epsilon_n (\bm{k})$ between energy $\epsilon$ and crystal momentum $\bm{k}$, with $n$ denoting a set of discrete indices such as band, spin, etc. Within this construction, the density of states (DOS) as a function of energy or momenta plays an important role in the calculation of physical properties such as heat capacity and magnetic susceptibility. While this is effectively a non-interacting picture, the electronic dispersion serves as an input for the treatment of interactions. In particular, an enhancement of the DOS at the Fermi level can strengthen the role of electron-electron interactions and lead to electronic instabilities.

In one and two dimensions, a divergence of DOS can occur due to the presence of one or more critical points in $\bm{k}$-space where $\nabla \epsilon (\bm{k}) = 0$~\cite{PhysRev.89.1189} (from here on we drop the index $n$). The simplest such singularity in two dimensions, the van Hove singularity (vHs), occurs due to regular saddle points and causes a logarithmic divergence of the DOS. The Hessian at a regular saddle point is non-degenerate [i.e., $\det (\partial_i \partial_j \, \epsilon) \neq 0$] and the dispersion  is quadratic to the lowest order [$\epsilon (\bm{k}) \sim k_x^2 - k_y^2$]. However there also exist higher order singularities which correspond to power law divergence of the DOS. These occur due to higher order critical points at which the Hessian is degenerate, i.e. non-invertible, and the dispersion needs to be expanded beyond quadratic order.

Singularities and the associated divergence of DOS are a signature of Fermi surface topological transitions~\cite{PhysRevB.95.035137,1810.13392} in which the Fermi surface undergoes a change in topology from electron type to hole type across the critical energy, with two or more branches of the Fermi surface touching at the critical point in a singular way.
Historically, when Lifshitz \cite{Lifshitz60} first studied the change of Fermi surface topology, he dealt with two cases: the appearance or collapsing of a neck and the appearance or collapsing of a pocket in the Fermi surface. 
The neck-collapsing case is the ordinary vHs, with the Fermi surface locally consisting of a pair of intersecting straight lines. These two types of Fermi surface topological transitions have been observed, along with their non-trivial consequences due to the presence of interactions, in a wide range of quantum materials including cuprates, iron arsenic and ferromagnetic superconductors, cobaltates, Sr$_2$RuO$_4$, heavy fermions \cite{Liu10,Okamoto10,HuxleyLifshitz11,Khan14,Benhabib15,Slizovskiy15,AokiCelrln16,Sherkunov18,Barber19}.

Higher order singularities display more exotic Fermi surface topological transitions. They have recently been associated with phenomena such as the unusual Landau level structure and tripling of de Haas$-$van Alphen and Shubnikov$-$de Haas oscillation periods in biased bilayer graphene~\cite{PhysRevB.95.035137}, the non-trivial thermodynamic and transport properties in $\mathrm{Sr_3 Ru_2 O_7}$~\cite{1810.13392}, correlated electron phenomena in twisted bilayer graphene near half filling~\cite{1901.05432} and the so called supermetal with diverging susceptibilities in the absence of long range order~\cite{1905.05188}. In the present work we develop a classification scheme for Fermi surface topological transitions and their associated DOS divergence using catastrophe theory. That catastrophe theory is an appropriate framework for examining higher order critical points in electronic bands as was first suggested by~\cite{PhysRevB.95.035137}. It has been also applied to other branches of physics (e.g. Ref (\onlinecite{Stewart82}), including optics \cite{Berry80} and molecular physics \cite{Kusmartsev89}).

Catastrophe theory deals with real valued functions of open subsets of $\mathbb{R}^n$, where $\mathbb{R}$  is the space of real numbers, and makes a distinction between state variables and control parameters. In the context of electronic bands the function of interest is the electronic dispersion and we identify the components of the crystal momentum as the state variables and hopping strengths, chemical potential, etc as control parameters which can be tuned externally, for example, by applying pressure, strain, bias voltage, etc. At a critical point, the gradient of the function with respect to the state variables vanishes. Since we focus on two dimensional systems, we are restricted to exactly two state variables, $k_x$ and $k_y$. With two state variables, catastrophe theory guarantees that a system with seven or less control parameters is typically equivalent to one of seventeen standard types of catastrophes, each of which corresponds to a unique higher order singularity. The higher order singularities are indexed by three positive integers: the \emph{corank, determinacy} and \emph{codimension}. The classification by these numbers is unique except for certain degenerate cases, which we show that in two-dimensions can be further distinguished by the \emph{winding}, i.e., the number of times the electronic dispersion changes sign along a contour encircling the critical point. \\

By tuning the control parameters, one can reach the higher order critical points corresponding to different catastrophes. In~\cite{1810.13392}, the fourth order saddle was identified with the unimodal parabolic singularity $X_9$ while in \cite{PhysRevB.95.035137} the monkey saddle was identified with the elliptic umbilic catastrophe. In the same spirit, we  identify the higher order singularity treated in~\cite{1901.05432} and~\cite{1905.05188} with the cusp catastrophe. \\

In the example in~\cite{1810.13392}, the elliptic umbilic catastrophe occurs on a point with $2\pi /3$ rotational symmetry, while in that of~\cite{1905.05188} the cusp occurs on a point with $\pi$ rotational symmetry. As a general principle, symmetry constrains a point to be critical and serve as the center where ordinary critical points merge. Only catastrophes consistent with the symmetry can occur at such a point. Another feature of high symmetry points in the Brillouin zone is that they can host otherwise atypical higher order singularities that are not part of the standard seventeen. These facts call for further classification of the catastrophes that can occur at the high symmetry points. We present such a classification for the Brillouin zones corresponding to the seventeen wallpaper groups (no relation to the seventeen catastrophes).\\

The paper is organized as follows: In Sec.~\ref{sec:ex1} we introduce the language of catastrophe theory through a simple example of a tight-binding model that displays a higher order singularity. This is followed by the classification of singularities in electronic bands in Sec.~\ref{sec:class}. The stability of this classification to small perturbations is discussed in Sec.~\ref{sec:robust}. In Sec.~\ref{sec:symmetries}, we explain the connection between high symmetry points in the Brillouin zone and higher order singularities. We then present a classification of the singularities allowed at the high symmetry points in the Brillouin zones corresponding to the seventeen wallpaper groups. In Sec.~\ref{sec:method}, we give a practical method for determining the type of singularity given its Taylor expansion and illustrate the method with a sample calculation. Finally we conclude the discussion in Sec.~\ref{sec:summary} by briefly sumarizing the scope of the work and setting the context for future work on the treatment of line singularities and the effects of interaction.

\section{A simple example}\label{sec:ex1}

In this section we introduce the languge of catastrophe theory through a simple tight binding model that displays a higher order singularity.\\

\begin{figure}[h]
 \centering
 \includegraphics[width=\columnwidth]{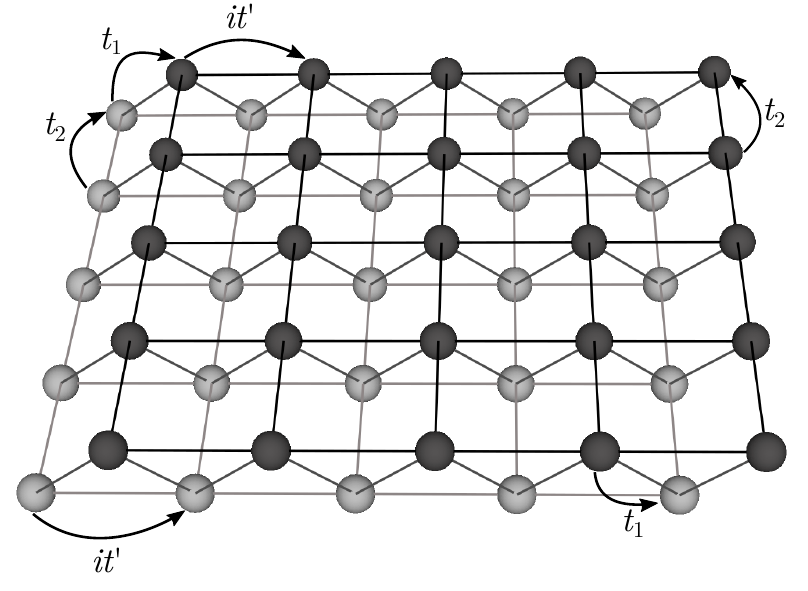}
 \caption{The lattice for the Hamiltonian treated in Sec~\ref{sec:simple_example} is shown above. It is a square lattice with a two site basis of atoms colored as black ($A$) and grey ($B$). The $AB$ nearest neighbor (NN) hopping in the $\hat{x}$ direction has a strength of $t_1$ while the $AA$ and $BB$ NN hoppings in the $\hat{y}$ direction have strength $t_2$. There is also complex next nearest neighbor $AA$ and $BB$ hopping $it'$ (with direction as depicted in the figure). These hopping strengths can be tuned so that the energy bands of this Hamiltonian yield the simplest of the higher order singularities, the fold.}
 \label{fig:lattice}
\end{figure}

The lattice pertaining to the model is layered with two sublattices $A$ and $B$, colored respectively by black and grey in Fig.~\ref{fig:lattice}. In the $\hat{y}$ direction, we have only $AA$ and $BB$ nearest neighbor (NN) hoppings of strength $t_2$. In the $\hat{x}$ direction, we have $A \rightarrow B$ NN hopping of strength $t_1$ and imaginary $A \rightarrow A$ and $B \rightarrow B$ next nearest neighbor (NNN) hopping $i t'$. The imaginary hopping can be interpreted as arising either from spin-orbit coupling or a staggered magnetic flux in the $\hat{y}$ direction.

\label{sec:simple_example}

Choosing a unit cell containing one $A$ atom and one $B$ atom as shown in Fig.~\ref{fig:lattice}, the Hamiltonian is given by:
\begin{multline}
H = \sum_{\bm{R}} \big[ t_1 \left( c_{A,\bm{R}}^{\dag} \, c_{B,\bm{R}} + \, c_{A,R}^{\dag} \, c_{B,\bm{R}-\hat{x}} \right) \\+ t_2 \left( c_{A,\bm{R}}^{\dag} \, c_{A,\bm{R}+\hat{y}} + \, c_{B,\bm{R}}^{\dag} \, c_{B,\bm{R}+\hat{y}} \right)\\ + it' \left( c_{A,\bm{R}+\hat{x}} \, c_{A,\bm{R}} + \, c_{B,\bm{R}+\hat{x}} \, c_{B,\bm{R}} \right) + \mathrm{h.c} \big],
\label{eq:toyham}
\end{multline}
where $c_{A/B,\bm{R}}$ annihilates an $A/B$ type Fermion in the unit cell located at $\bm{R}$. Diagonalizing this Hamiltonian in $k$-space yields two bands: $\epsilon_{\pm}(\bm{k}) = \pm 2 t_1 \, \cos \, (k_x/2) - 2t' \, \sin \, (k_x) + 2t_2 \, \cos \, (k_y)$. We now expand $\epsilon_+(\bm{k})$ to cubic order around the point $(\pi,0)$ in the first Brillouin Zone (BZ):
 \begin{equation}
 \epsilon_+(\bm{k}) \approx 2t_2 + (2t' - t_1) k_x + ( t_1 /4 - 2t') k_x^3 - t_2 \, k_y^2,
 \end{equation}
where $\alpha = 2t'-t_1$ and $\beta^3 = t_1/4-2t'$ are independent. To obtain a critical point, we need to remove the $k_x$ term by tuning $\alpha$ to zero. We assume $\beta \neq 0$ and $t_2 \neq 0$. This guarantees that we can always rescale $(k_x,k_y) \rightarrow (k_x/\beta, k_y/\sqrt{t_2})$ and write $t=\alpha / \beta$ to obtain: 
\begin{equation}
\epsilon_+ (\bm{k}) = 2t_2 + k_x^3 - k_y^2 + t k_x
\label{eq:unfolding1}
\end{equation}
Thus, up to a rescaling of coordinates and a constant energy shift ($2t_2$), we effectively have just one control parameter $t$ in this dispersion. \\

\begin{figure*}[!ht]
 \centering
 \includegraphics[width=1\textwidth]{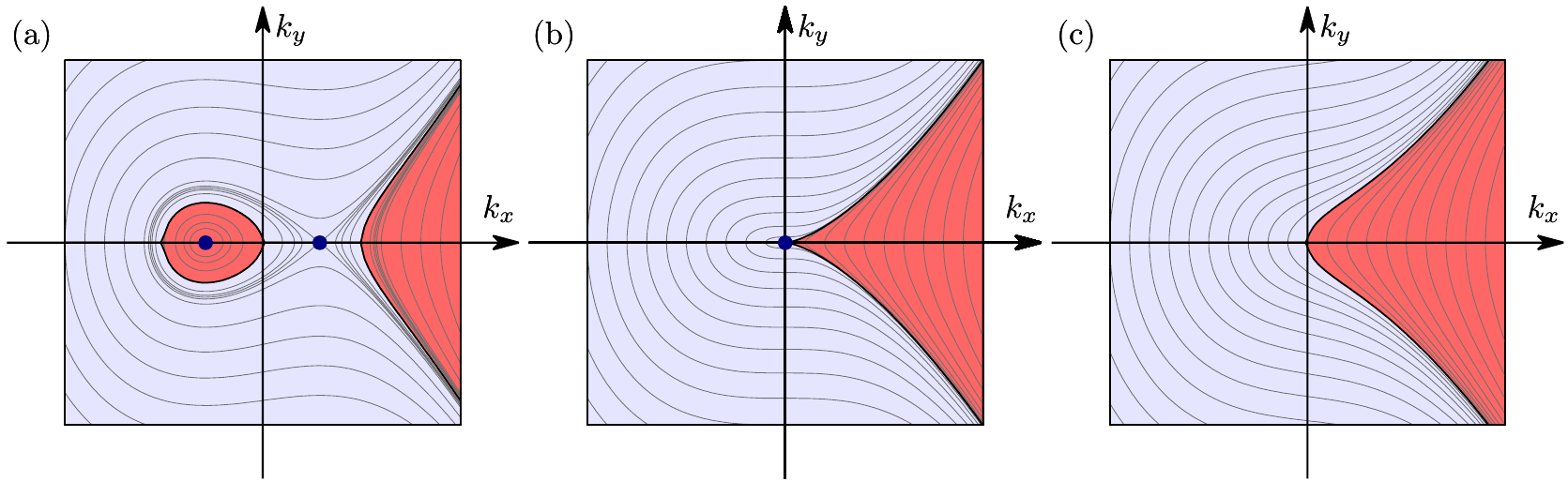}
 \caption{Constant energy contours of the dispersion $f(k_x, k_y) = k_x^3 - k_y^2 + t k_x$ for three cases: (a) $t < 0$, (b) $t = 0$ and (c) $t > 0$. When $t < 0$ there is an ordinary maximum and an ordinary minimum. As $t$ is reduced to $0$, these merge at the origin to give a higher order singularity. For $t > 0$ there is no longer a critical point.}
 \label{fig:example}
\end{figure*}

In Fig.~\ref{fig:example} we illustrate the behaviour of the energy contours and the Fermi surface for three distinct cases. When $t<0$, the Fermi sea is topologically non-trivially connected, with an island of positive energies (small red region) in the middle of the Fermi sea (blue region). There are two ordinary critical points in $k$-space, demarked with thick dots: a maximum in the dispersion (the summit of the red island) and minimum (bottom of the blue Fermi sea). Upon increasing $t$ to 0, a higher order critical point ($k_x^3-k_y^2$) appears suddenly. This higher order critical point at $t=0$ corresponds to the merger of the two ordinary critical points seen for $t<0$, as well as to the Fermi surface topological transition in which the island disappears and the Fermi sea becomes trivially connected. When $t>0$, the dispersion is no longer singular, and the single component Fermi sea persists.

At the higher order critical point ($t=0$), the DOS diverges as $|\epsilon - 2t_2|^{-1/6}$ when $\epsilon\to 2t_2$. By including higher order terms in the dispersion we still obtain the same power law divergence to the lowest order and the Fermi surface topology does not change. The value of the exponent is a property of the type of catastrophe, and cannot be changed by smooth coordinate transformations, as we show in Appendix~\ref{app:DOS}. For the singularity in this simple example, the cubic truncation of the Taylor series is sufficient.

As the above example shows, we can often tune some control parameters in a system to obtain a higher order critical point. By a suitable smooth transformation of the coordinates and the control parameters, we can put the dispersion in a convenient form that contains the higher order singularity, to which polynomial terms modulated by the control parameters are added (Eq.~\ref{eq:unfolding1}). For a given higher order singularity, there is usually a most general such expression, known as the \emph{universal unfolding}, containing only a finite number of effective control parameters modulating polynomial perturbations ($t\, k_x$ in the example). These polynomial perturbations represent directions in polynomial space which are not tangent to the orbit of the higher order singularity under smooth coordinate transformations. The number of such missing directions (or equivalently effective control parameters) is the \emph{codimension} of the singularity. Finally, these higher order singularities are well described by truncating the Taylor expansion to a certain lowest order, known as the \emph{determinacy}. In the example above, the determinancy is 3, justifying the truncation to cubic order.

In addition to the codimension and determinacy, catastrophy theory utilizes another index, the \emph{corank}, which is the number of zero eigenvalues of the Hessian matrix of the function at the critical point. (See Appendix~\ref{app:catastrophe_review} for a self-contained review of catastrophe theory). While these three indices can classify singularities in any dimension, they are not sufficient to resolve the degeneracies. In the particular case of functions of two variables, (for instance $k_x$ and $k_y$ in the case of dispersions of two-dimensional electronic systems), we find that these degeneracies can be resolved by introducing a \emph{fourth} index, the \emph{winding} number, which counts the number of times the function changes sign in a small loop around the origin.

In terms of these four indices (codimension, determinacy, corank, and winding), we can enumerate all possible singularities that the dispersion of a two-dimensional electronic system would typically present. These singularities, along with the associated power laws in the DOS and universal ratios of their prefactors, are listed in the next section.

\section{Classification of higher order singularities in electronic systems}\label{sec:class}

Using catastrophe theory we can classify all possible higher order
singularities that can occur in the electronic dispersion of
two-dimensional electrons. If there are seven or less control
parameters in the system, catastrophe theory guarantees that only
higher order singularities with codimension (or cod) $\leqslant 7$ are
typically likely to occur. Typicality here has a simple but precise
mathematical meaning, that is best illustrated by an example. Typically
a set of three equations in two variables is over determined and does
not have a solution (two straight lines in three dimensional space
do not typically intersect), while a system of two
equations in two variables typically has a unique solution, at least
locally (two straight lines in two dimension typically intersect at a
point).

There are 17 singularities with cod $\leqslant 7$, which we list in
Tables~\ref{table1},~\ref{table2} and~\ref{table3}. We also include one other atypical singularity,
the $X_9$, which has codimension $8$. It would normally require 8 control parameters, but
crystal symmetry ensures that some of the constraints are automatically
satisfied. In fact it is the lowest codimension singularity that respects four-fold rotational symmetry. The tables are organized by the four integer indices (cor,
cod, det, w) corresponding to corank, codimension, determinacy, and
winding.

As mentioned earlier, at an ordinary critical point in two dimensions, the DOS has a logarithmic divergence. At higher order critical points, the DOS diverges in a power law fashion, often with different coefficients $D_+$ and $D_-$ as we approach the critical energy from above and below:  
\begin{equation}
g(\epsilon) = \begin{cases}
D_+ |\epsilon|^{-\gamma}, &\epsilon > 0\\
D_- |\epsilon|^{-\gamma}, &\epsilon < 0
\end{cases}
\;.
\end{equation}
In Appendix~\ref{app:DOS} we show that both the exponent and the ratio of the coefficients are universal in
that they are preserved under smooth coordinate transformations. The power law
dependence for dispersions with two monomial terms can easily be
extracted by scaling $k_x$ and $k_y$ appropriately, as mentioned
in~\cite{1905.05188}. But this procedure by itself does not give the
prefactors and one would still have to convert the DOS integral into
an integral over constant energy contours to evaluate $D_{\pm}$. This
is also done in Appendix~\ref{app:DOS}. In the tables, we list for
each catastrophe both the exponent $\gamma$ controlling the divergence
of the DOS at the critical point and the universal ratio
$D_+/D_-$.



\begin{table*}
\begin{center}
\begin{tabular}{|c|c|c|c|c|c|}
\hline
\begin{tabular}{@{}c@{}}
Catastrophe\\
(ADE index)
\end{tabular}
&
(cor, cod, det, w)
&
DOS
&
$D_+/D_-$
&
\begin{tabular}{@{}c@{}}
Universal\\
unfolding
\end{tabular}
&
\begin{tabular}{@{}c@{}}
Energy contours\\
(for $t_i = 0$)
\end{tabular}\\
\hline
\begin{tabular}{@{}c@{}}
Fold\\
($A_2$)
\end{tabular}
&
$(1,1,3,2)$
&
$|\epsilon|^{-1/6}$
&
$\frac{1}{\sqrt{3}}$
&
\begin{tabular}{@{}c@{}}
$k_x^3-k_y^2 + t_1 k_x$
\end{tabular}
&
\begin{tabular}{@{}c@{}}
\includegraphics[scale=1]{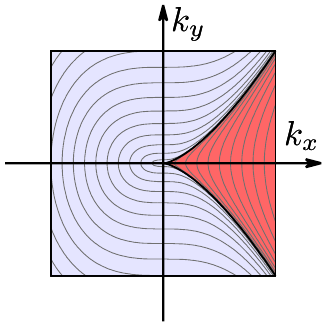}
\end{tabular}\\
\begin{tabular}{@{}c@{}}
Cusp\\
($A_3$)
\end{tabular}
&
$(1,2,4,4)$
&
$|\epsilon|^{-1/4}$
&
$\frac{1}{\sqrt{2}}$
&
\begin{tabular}{@{}c@{}}
$k_x^4-k_y^2 + t_2 k_x^2 + t_1 k_x$
\end{tabular}
&
\begin{tabular}{@{}c@{}}
\includegraphics[scale=1]{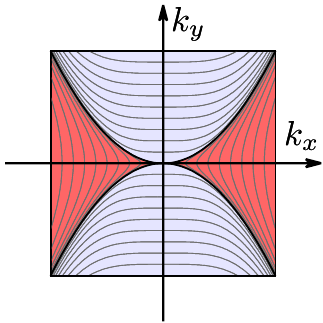}
\end{tabular}\\
\begin{tabular}{@{}c@{}}
Swallowtail\\
($A_4$)
\end{tabular}
&
$(1,3,5,2)$
&
$|\epsilon|^{-3/10}$
&
$\sqrt{1-\frac{2}{\sqrt{5}}}$
&
\begin{tabular}{@{}c@{}}
$k_x^5-k_y^2 +t_3 k_x^3 + t_2 k_x^2 + t_1 k_x$
\end{tabular}
&
\begin{tabular}{@{}c@{}}
\includegraphics[scale=1]{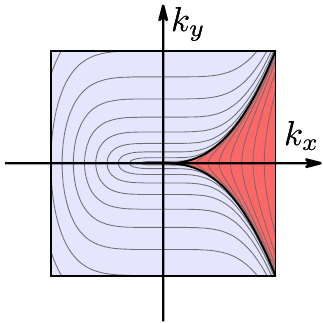}
\end{tabular}\\
\begin{tabular}{@{}c@{}}
Butterfly\\
($A_5$)
\end{tabular}
&
$(1,4,6,4)$
&
$|\epsilon|^{-1/3}$
&
$\frac{1}{2}$
&
\begin{tabular}{@{}c@{}}
$k_x^6-k_y^2 + t_4 k_x^4 + t_3 k_x^3 + t_2 k_x^2 + t_1 k_x $
\end{tabular}
&
\begin{tabular}{@{}c@{}}
\includegraphics[scale=1]{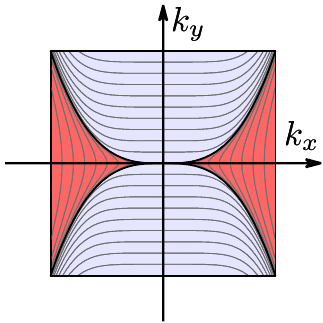}
\end{tabular}\\
\begin{tabular}{@{}c@{}}
Wigwam\\
($A_6$)
\end{tabular}
&
$(1,5,7,2)$
&
$|\epsilon|^{-5/14}$
&
$\tan \left(\frac{\pi }{14}\right)$
&
\begin{tabular}{@{}c@{}}
$k_x^7-k_y^2 + t_5 k_x^5 + t_4 k_x^4 + t_3 k_x^3 + t_2 k_x^2 + t_1 k_x$
\end{tabular}
&
\begin{tabular}{@{}c@{}}
\includegraphics[scale=1]{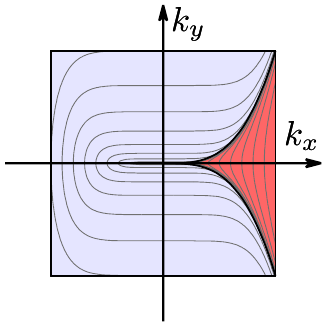}
\end{tabular}\\
\begin{tabular}{@{}c@{}}
Star\\
($A_7$)
\end{tabular}
&
$(1,6,8,4)$
&
$|\epsilon|^{-3/8}$
&
$\sin \left(\frac{\pi }{8}\right)$
&
\begin{tabular}{@{}c@{}}
$k_x^8-k_y^2 + t_6 k_x^6 + t_5 k_x^5 + t_4 k_x^4 + t_3 k_x^3 + t_2 k_x^2 + t_1 k_x $
\end{tabular}
&
\begin{tabular}{@{}c@{}}
\includegraphics[scale=1]{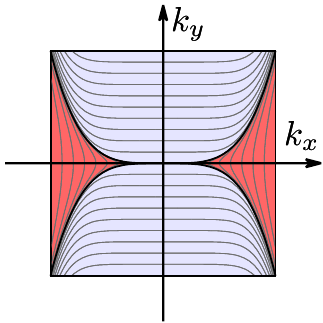}
\end{tabular}\\
\hline
\end{tabular}
\end{center}
\caption{The standard catastrophes are uniquely indexed by corank (cor), codimension (cod), determinacy (det) and winding (w). The DOS diverges in a power law fashion with a universal ratio ($D_+/D_-$) for the coefficients. The cusoid catastrophes $A_{n}$, tabulated above, have cor $=1$ and a dispersion of the form $k_x^{n+1} - k_y^2$ up to an overall sign. (We do not consider cuspoids of the form $\pm(k_x^{2m}+k_y^2)$ since above/below the extremum the Fermi surface ceases to exist). The universal unfolding contains the directions missing from the orbit of the singularity under smooth coordinate transformations. The number of these directions is equal to the codimension.}
\label{table1}
\end{table*}



\begin{table*}
\begin{center}
\begin{tabular}{|c|c|c|c|c|c|}
\hline
\begin{tabular}{@{}c@{}}
Catastrophe\\
(ADE index)
\end{tabular}
&
(cor, cod, det, w)
&
DOS
&
$D_+/D_-$
&
\begin{tabular}{@{}c@{}}
Universal\\
unfolding
\end{tabular}
&
\begin{tabular}{@{}c@{}}
Energy contours\\
(for $t_i = 0$)
\end{tabular}\\
\hline
\begin{tabular}{@{}c@{}}
Elliptic\\
umbilic\\
($D_4^-$)
\end{tabular}
&
$(2,3,3,6)$
&
$|\epsilon|^{-1/3}$
&
$1$
&
\begin{tabular}{@{}c@{}}
$3k_x^2 k_y-k_y^3 +t_3 (k_x^2 + k_y^2)+t_2 k_y +t_1 k_x$
\end{tabular}
&
\begin{tabular}{@{}c@{}}
\includegraphics[scale=1]{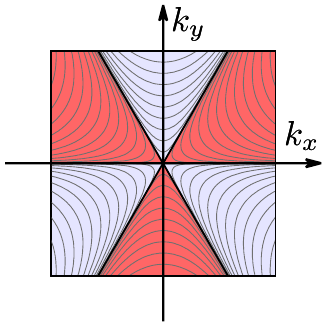}
\end{tabular}\\
\begin{tabular}{@{}c@{}}
Hyperbolic\\
umbilic\\
($D_4^+$)
\end{tabular}
&
$(2,3,3,2)$
&
$|\epsilon|^{-1/3}$
&
$1$
&
\begin{tabular}{@{}c@{}}
$k_x^3+k_y^3 +t_3 k_x k_y +t_2 k_x + t_1 k_y $
\end{tabular}
&
\begin{tabular}{@{}c@{}}
\includegraphics[scale=1]{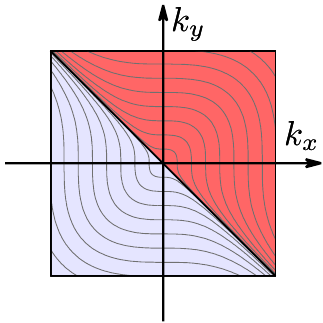}
\end{tabular}\\
\begin{tabular}{@{}c@{}}
Parabolic\\
umbilic\\
($D_5$)
\end{tabular}
&
$(2,4,4,4)$
&
$|\epsilon|^{-3/8}$
&
$\frac{1}{2} \csc \left(\frac{\pi }{8}\right)$
&
\begin{tabular}{@{}c@{}}
$k_x^2 k_y+k_y^4 + t_4 k_x^2 + t_3 k_y^2 + t_2 k_x + t_1 k_y$
\end{tabular}
&
\begin{tabular}{@{}c@{}}
\includegraphics[scale=1]{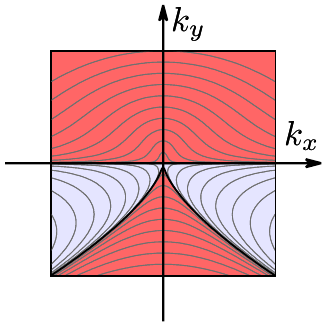}
\end{tabular}\\
\begin{tabular}{@{}c@{}}
Symbolic\\
umbilic\\
($E_6$)
\end{tabular}
&
$(2,5,4,2)$
&
$|\epsilon|^{-5/12}$
&
$\sqrt{2+\sqrt{3}}$
&
\begin{tabular}{@{}c@{}}
$k_x^4+k_y^3 + t_5 k_x^2 k_y + t_4 k_x^2 + t_3 k_x k_y  + t_2 k_x + t_1 k_y $
\end{tabular}
&
\begin{tabular}{@{}c@{}}
\includegraphics[scale=1]{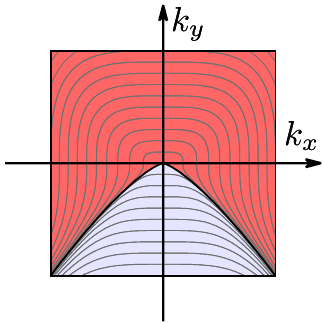}
\end{tabular}\\
\begin{tabular}{@{}c@{}}
Second\\
elliptic\\
umbilic\\
($D_6^-$)
\end{tabular}
&
$(2,5,5,6)$
&
$|\epsilon|^{-2/5}$
&
$1$
&
\begin{tabular}{@{}c@{}}
$k_x^2 k_y-k_y^5 +t_5 k_y^4 + t_4 k_y^3 + t_3 k_y^2 + t_2 k_y + t_1 k_x$
\end{tabular}
&
\begin{tabular}{@{}c@{}}
\includegraphics[scale=1]{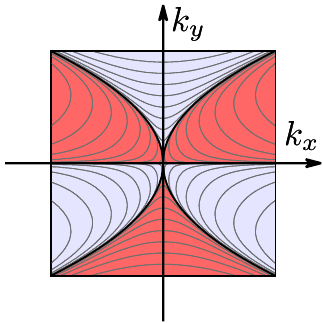}
\end{tabular}\\
\begin{tabular}{@{}c@{}}
Second\\
hyperbolic\\
umbilic\\
($D_6^+$)
\end{tabular}
&
$(2,5,5,2)$
&
$|\epsilon|^{-2/5}$
&
$1$
&
\begin{tabular}{@{}c@{}}
$k_x^2 k_y+k_y^5 +t_5 k_y^4 + t_4 k_y^3 + t_3 k_y^2 + t_2 k_y + t_1 k_x$
\end{tabular}
&
\begin{tabular}{@{}c@{}}
\includegraphics[scale=1]{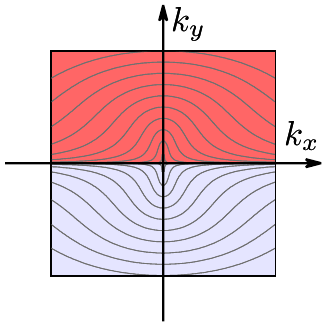}
\end{tabular}\\
\hline
\end{tabular}
\end{center}
\caption{The umbilic catastrophes tabulated above have corank two. Like the cuspoids listed in Table~\ref{table1} they have saddle like nature with a non-zero winding number number that counts the number of times the dispersion changes sign along a closed contour enclosing the origin. However, unlike the cuspoids wherein those with even degree have $\pi$ rotation symmetry, most of the umbilics do not possess any rotational symmetry. The elliptic umbilic (monkey saddle) is an exception in that it has $2\pi/3$ rotation symmetry.}
\label{table2}
\end{table*}



\begin{table*}
\begin{center}
\begin{tabular}{|c|c|c|c|c|c|}
\hline
\begin{tabular}{@{}c@{}}
Catastrophe\\
(ADE index)
\end{tabular}
&
(cor, cod, det, w)
&
DOS
&
$D_+/D_-$
&
\begin{tabular}{@{}c@{}}
Universal\\
unfolding
\end{tabular}
&
\begin{tabular}{@{}c@{}}
Energy contours\\
(for $t_i = 0$)
\end{tabular}\\
\hline
\begin{tabular}{@{}c@{}}
$E_7$
\end{tabular}
&
$(2,6,4,4)$
&
$|\epsilon|^{-4/9}$
&
$1$
&
\begin{tabular}{@{}c@{}}
$k_x^3 k_y+k_y^3 + t_6 k_x k_y^2 +t_5 k_x k_y + t_4 k_x^2 + t_3 k_y^2 $\\$ +t_2 k_x +t_1 k_y$
\end{tabular}
&
\begin{tabular}{@{}c@{}}
\includegraphics[scale=1]{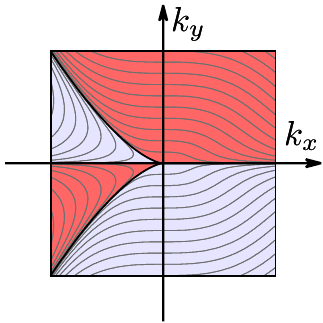}
\end{tabular}\\
\begin{tabular}{@{}c@{}}
$D_7$
\end{tabular}
&
$(2,6,6,4)$
&
$|\epsilon|^{-5/12}$
&
$\sqrt{\frac{3}{2}}$
&
\begin{tabular}{@{}c@{}}
$k_x^2 k_y+k_y^6 + t_6 k_y^5 +t_5 k_y^4 + t_4 k_y^3 + t_3 k_y^2 +t_2 k_y $\\$ +t_1 k_x$
\end{tabular}
&
\begin{tabular}{@{}c@{}}
\includegraphics[scale=1]{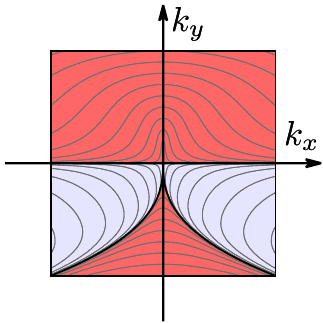}
\end{tabular}\\
\begin{tabular}{@{}c@{}}
$E_8$
\end{tabular}
&
$(2,7,5,2)$
&
$|\epsilon|^{-7/15}$
&
$1$
&
\begin{tabular}{@{}c@{}}
$k_x^3 + k_y^5  + t_7 k_x k_y^3 + t_6 k_x k_y^2 + t_5 k_y^3  + t_4 k_x k_y $\\$ + t_3 k_y^2 + t_2 y + t_1 k_x$
\end{tabular}
&
\begin{tabular}{@{}c@{}}
\includegraphics[scale=1]{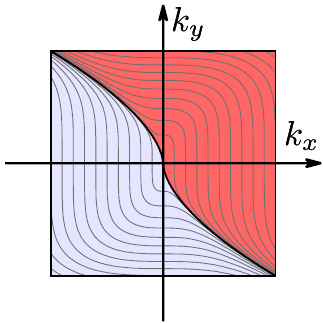}
\end{tabular}\\
\begin{tabular}{@{}c@{}}
$D_8^+$
\end{tabular}
&
$(2,7,7,2)$
&
$|\epsilon|^{-3/7}$
&
$1$
&
\begin{tabular}{@{}c@{}}
$k_x^2 k_y+k_y^7 + t_7 k_y^6 + t_6 k_y^5 + t_5 k_y^4  + t_4 k_y^3 + t_3 k_y^2 $\\$ + t_2 k_y $
\end{tabular}
&
\begin{tabular}{@{}c@{}}
\includegraphics[scale=1]{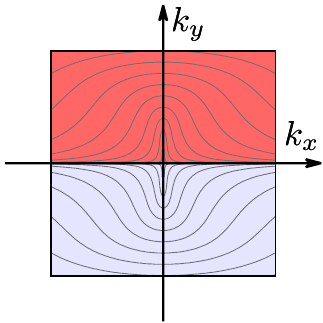}
\end{tabular}\\
\begin{tabular}{@{}c@{}}
$D_8^-$
\end{tabular}
&
$(2,7,7,6)$
&
$|\epsilon|^{-3/7}$
&
$1$
&
\begin{tabular}{@{}c@{}}
$k_x^2 k_y-k_y^7 + t_7 k_y^6 + t_6 k_y^5 + t_5 k_y^4  + t_4 k_y^3 + t_3 k_y^2 $\\$ + t_2 k_y +t_1 k_x $
\end{tabular}
&
\begin{tabular}{@{}c@{}}
\includegraphics[scale=1]{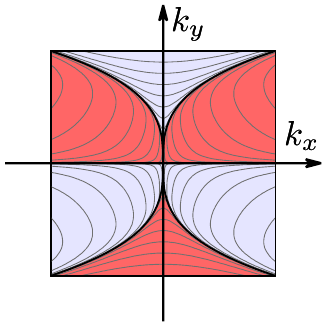}
\end{tabular}\\
\begin{tabular}{@{}c@{}}
$X_9$
\end{tabular}
&
\begin{tabular}{@{}c@{}}
$(2,8,4,8)$\\
(for $c = -3$)
\end{tabular}
&
$|\epsilon|^{-1/2}$
&
\begin{tabular}{@{}c@{}}
$1$\\
(for $c = -3$)
\end{tabular}
&
\begin{tabular}{@{}c@{}}
$k_x^4 + k_y^4 +2ck_x^2 k_y^2 + t (k_x^2 + k_y^2)$\\
(unfolding consistent with $\pi/2$-rotation).
\end{tabular}
&
\begin{tabular}{@{}c@{}}
\includegraphics[scale=1]{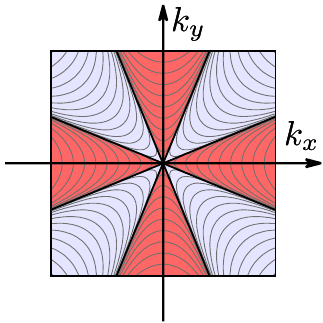}
\end{tabular}\\
\hline
\end{tabular}
\end{center}
\caption{The original Thom's theorem dealt with seven catastrophes. It was first extended to twelve and then to seventeen catastrophes. The last five catastrophes are documented above along with the $X_9$ singularity which is the singularity of lowest codimension that is allowed by $\pi / 2$ rotation. Unlike the rest, it is unimodal: different values of the parameter $c$ in the dispersion correspond in general to inequivalent singularities with different windings. The Fermi surface shown above is for $c = -3$, a generalization of the monkey saddle.}
\label{table3}
\end{table*}


\section{Robustness of the classification}\label{sec:robust}
Given that higher order critical points are somewhat rare and may require tuning of parameters to obtain, it is natural to ask how robust the classification scheme presented above is. This is particularly relevant in real systems which are prone to imperfections manifesting as small perturbations, motivating the question: can small perturbations change the type of higher order singularity that can occur on tuning parameters? The answer is no. This is because the universal unfolding contains \emph{all} possible perturbations to the higher order singularity. For any sufficiently small perturbation, we get the same catastrophe, up to a smooth coordinate transformation. Only the critical value of the tuning parameters for which we obtain the higher order singularity, gets shifted. This property is referred to as stability in the literature. It is important to note that stability is guaranteed only for singularities with codimension $\leqslant 7$, viz the seventeen listed above. 

For codimension 8 and higher we encounter the so called unimodal singularities which possess a continuous parameter in their dispersion. Two singularities with different values of the continuous parameter, however close by, will in general not be smoothly equivalent so that a small perturbation can take us to an altogether different singularity. But as mentioned earlier these are atypical in systems where seven or less parameters are tuned.


\section{High symmetry points and catastrophes}
\label{sec:symmetries}

As seen in Sections~\ref{sec:simple_example} and~\ref{sec:class}, within a Bloch-band description, a point in $k$-space  can morph into a higher order critical point when the parameters in the Hamiltonian are tuned. While this is may be hard to achieve at an arbitrary point, high symmetry points in the Brillouin zone are particularly easy to tune into higher order critical points. Recall that for a point to be critical, we need both components of the Jacobian at the point to vanish. For higher order critical points, we further need the determinant of the Hessian to vanish (or equivalently at least one of the eigenvalues needs to be zero). If a reflection axis passes through the point, the component of the Jacobian perpendicular to the axis is already constrained to be zero, yielding a simplification. Alternatively, if the point is a center of a non-trivial rotation, the entire Jacobian will vanish so that we have to tune only the determinant of the Hessian to zero (see Appendix~\ref{app:symmetry} for details). Thus it is relatively easier to obtain higher order singularities at points of high symmetry by tuning parameters in the Hamiltonian.
\begin{figure}[h]
 \centering
 \includegraphics[width=\columnwidth]{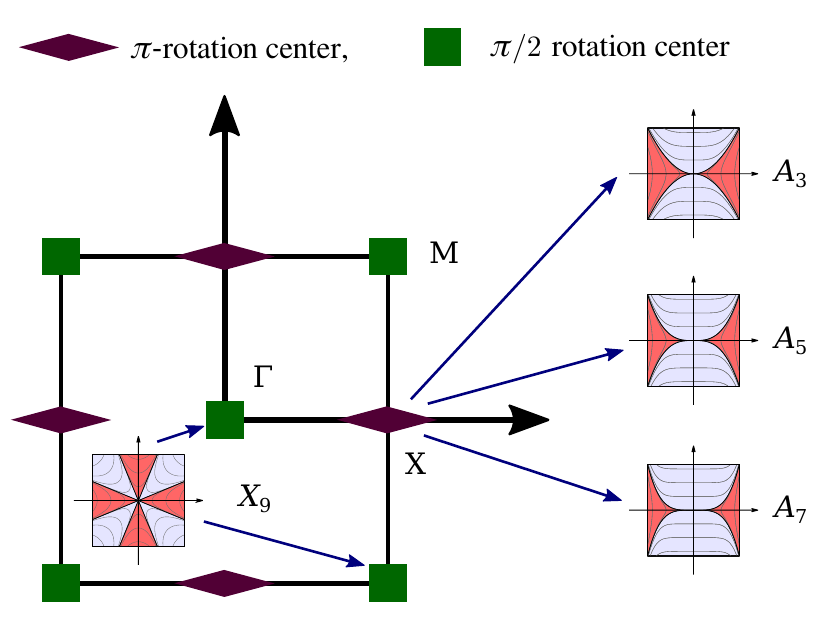}
 \caption{The Brillouin zone of the wallpaper group p4 shown along with the high symmetry points. The point group of p4 is generated by $\pi/2$-rotation which gets carried over to $k$-space as a symmetry that acts about the origin. By combining the $\pi/2$ and $\pi$ rotations with reciprocal lattice translations we can identify the other high symmetry points. At each of the high symmetry points, only dispersions which respect the symmetry are allowed. This places a huge restriction on the higher order singularities that are likely to occur. In the insets we identify the catastrophes likely to occur at each of the high symmetry points.}
 \label{fig:wallpaper_example}
\end{figure}

Having stressed the importance of high symmetry points in $k$-space, we remark that they can be identified in a systematic way (see for instance~\citep{dresselhaus2007group}). Given a lattice with a certain symmetry group, the dispersing bands have the symmetry of the point group. If the Hamiltonian is time reversal invariant (TRI) and there is no spin orbit coupling, we have an additional inversion symmetry in $k$-space. By combining the point group symmetries with reciprocal lattice translations we can find the high symmetry points in the first BZ. We illustrate this for the wallpaper group p4 in Fig~\ref{fig:wallpaper_example}. The point group of p4 is $C_4$, which is generated by $\pi/2$-rotation. This point group gets carried to $k$-space wherein it acts about the $\Gamma$ point (the origin in $k$-space). $\Gamma$ is thus a center of four fold rotation. We then combine the $\pi/2$-rotation and $\pi$-rotation with various reciprocal lattice translations. Combining rotations with translations simply shits the center of rotation. This procedure then yields the $M$ and $X$ points as centers of 4-fold and 2-fold rotations respectively in the first BZ.\\

At the high symmetry points, only singularities consistent with the symmetry of the point are allowed. For example, at the $\Gamma$ and $M$ points in the above example, only singularities which are invariant under $\pi/2$ rotation are allowed to occur. $X_9$ is the singularity with lowest codimension that satisfies this property and is thus the one that is typically likely to occur. At the $X$ point, only the cuspoids $A_{2n-1}$ with the standard form $k_x^{2n} \pm k_y^2$ for $n=1,2,3$ are allowed to occur. This is because these are the only ones among the seventeen which are invariant under $\pi$ rotation. (It can be shown that no smooth coordinate transformation can make the rest among the seventeen consistent with $\pi$ rotation). In Table~\ref{table4} we apply this scheme for the Brillouin zones corresponding to the seventeen wallpaper groups in two dimension. (See Appendix~\ref{app:isometries} for a quick review of crystal symmetries and their consequences for bands).\\


\begin{table*}
\begin{center}

\begin{tabular}{|cc|cc|c|}
\hline
\begin{tabular}{@{}c@{}}
\includegraphics{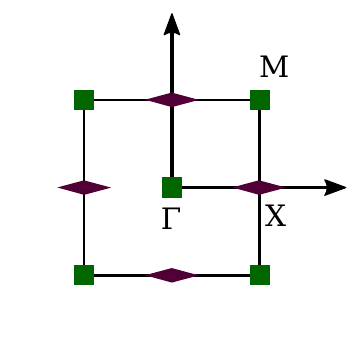}
\end{tabular}
& \begin{tabular}{@{}c@{}}
\textbf{p4} ($\mathbf{C_4}$)\\
(Square)\\ \\
$\Gamma$: $X_9$\\
X: $A_{2n-1}$\\
M: $X_9$
\end{tabular} & 
\begin{tabular}{@{}c@{}}
\includegraphics{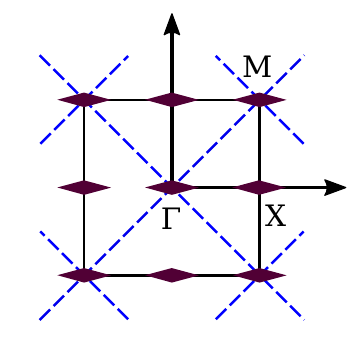}
\end{tabular}
&\begin{tabular}{@{}c@{}}
\textbf{cmm} ($\mathbf{C_2}$)\\
(Square)\\ \\
$\Gamma$: $A_{2n-1}$\\
X: $A_{2n-1}$\\
M: $A_{2n-1}$
\end{tabular}
&
\begin{tabular}{cc}
\begin{tabular}{@{}c@{}}
\includegraphics{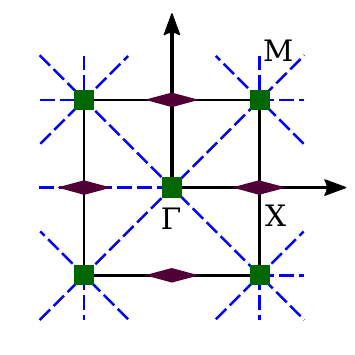}
\end{tabular}
& \begin{tabular}{@{}c@{}}
\textbf{p4m, p4g} ($\mathbf{D_4}$)\\ 
(Square)\\ \\
$\Gamma$: $X_9$\\
X: $A_{2n-1}$\\
M: $X_9$
\end{tabular}
\end{tabular}
\\
\hline
\begin{tabular}{@{}c@{}}
\includegraphics{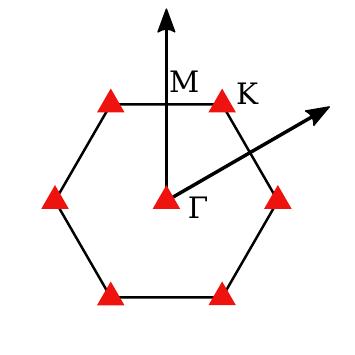}
\end{tabular}&
\begin{tabular}{@{}c@{}}
\textbf{p3} ($\mathbf{C_3}$)\\
(Triangular)\\\\
$\Gamma$: $D_4^-$\\
M: All\\
K: $D_4^-$
\end{tabular}&
\begin{tabular}{@{}c@{}}
\includegraphics{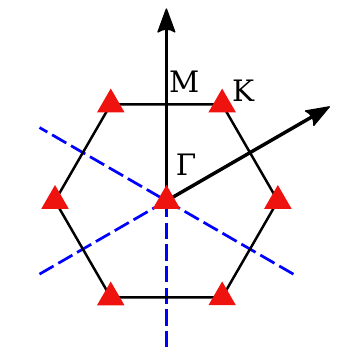}
\end{tabular}&
\begin{tabular}{@{}c@{}}
\textbf{p3m1} ($\mathbf{D_3}$)\\
(Triangular)\\\\
$\Gamma$: $D_4^-$\\
M: All\\
K: $D_4^-$
\end{tabular}
&
\begin{tabular}{cc}
\begin{tabular}{@{}c@{}}
\includegraphics{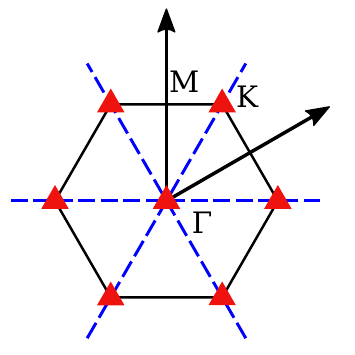}
\end{tabular}
& 
\begin{tabular}{@{}c@{}}
\textbf{p31m} ($\bf{D_3}$)\\
(Triangular)\\\\
$\Gamma$: $D_4^-$\\
M: All\\
K: $D_4^-$
\end{tabular}
\end{tabular}
\\
\hline
\begin{tabular}{@{}c@{}}
\includegraphics{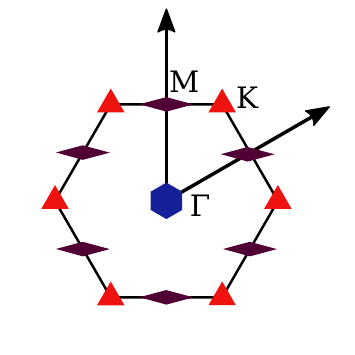} 
\end{tabular} &
\begin{tabular}{@{}c@{}}
\textbf{p6} ($\mathbf{C_6}$), \textbf{p3*}\\
(Triangular)\\\\
$\Gamma$: None\\
M: $A_{2n-1}$\\
K: $D_4^-$
\end{tabular} &
\begin{tabular}{@{}c@{}}
\includegraphics{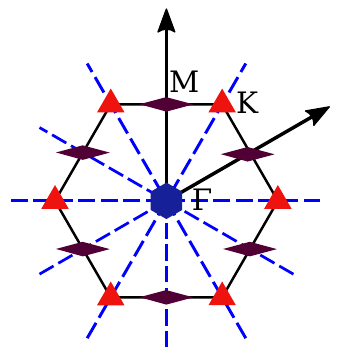}
\end{tabular} &
\begin{tabular}{@{}c@{}}
\textbf{p6m} ($\mathbf{D_6}$),\\
\textbf{p31m*}, \textbf{p3m1*}\\
(Triangular)\\\\
$\Gamma : \,$None \\
M: $A_{2n-1}$\\
K: $D_4^-$
\end{tabular} &
\begin{tabular}{cc}
\begin{tabular}{@{}c@{}}
\includegraphics{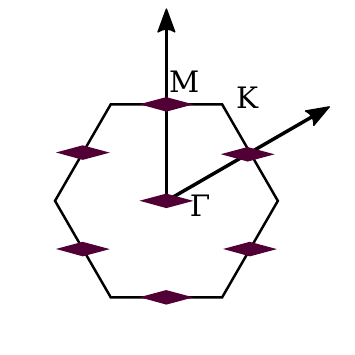}
\end{tabular}
&
\begin{tabular}{@{}c@{}}
\textbf{p2} ($\mathbf{C_2}$), \textbf{p1*}\\
(Triangular)\\\\
$\Gamma$: $A_{2n-1}$\\
M: $A_{2n-1}$\\
K: All
\end{tabular}
\end{tabular}
\\
\hline
\begin{tabular}{@{}c@{}}
\includegraphics{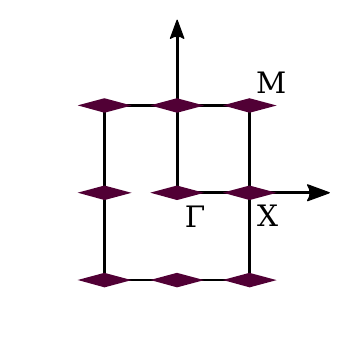}
\end{tabular}
&
\begin{tabular}{@{}c@{}}
\textbf{p2} ($\mathbf{C_2}$), \textbf{p1*}\\
(Rectangular)\\\\
$\Gamma$: $A_{2n-1}$\\
X: $A_{2n-1}$\\
M: $A_{2n-1}$
\end{tabular}
&
\begin{tabular}{@{}c@{}}
\includegraphics{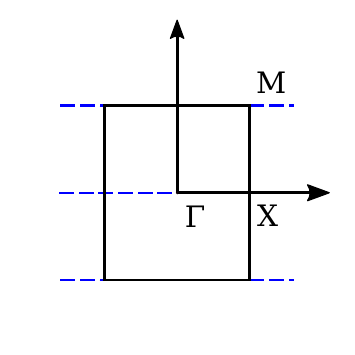}
\end{tabular} &
\begin{tabular}{@{}c@{}}
\textbf{pm, pg} ($\mathbf{D_1}$) \\
(Rectangular)\\\\
$\Gamma$: All\\
X: All\\
M: All
\end{tabular} &
\begin{tabular}{cc}
\begin{tabular}{@{}c@{}}
\includegraphics{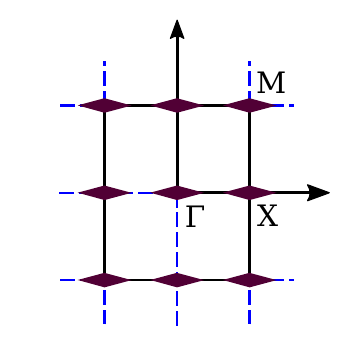} 
\end{tabular} &
\begin{tabular}{@{}c@{}}
\textbf{pmm, pmg}\\ \textbf{pgg} ($\mathbf{D_2}$), \\
\textbf{pm*}, \textbf{pg*}\\
(Rectangular\\
\& Square)\\\\
$\Gamma$: $A_{2n-1}$\\
X: $A_{2n-1}$\\
M: $A_{2n-1}$
\end{tabular} 
\end{tabular}
\\
\hline
\begin{tabular}{@{}c@{}}
\includegraphics{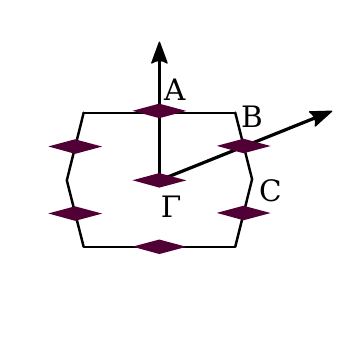} 
\end{tabular} &
\begin{tabular}{@{}c@{}}
\textbf{p2} ($\mathbf{C_2}$), \textbf{p1*} \\ (Rhombohedral) \\\\
$\Gamma$: $A_{2n-1}$\\
A: $A_{2n-1}$\\
B: $A_{2n-1}$\\
C: $A_{2n-1}$
\end{tabular} 
&
\begin{tabular}{@{}c@{}}
\includegraphics{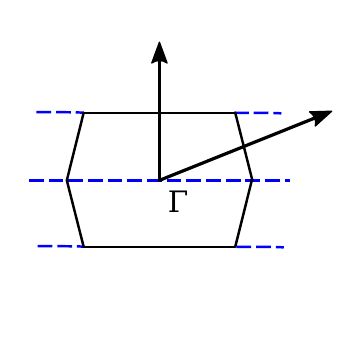} 
\end{tabular} &
\begin{tabular}{@{}c@{}}
\textbf{cm} ($\mathbf{D_1}$) \\ (Rhombohedral) \\\\
$\Gamma$: All\\
\end{tabular} 
&
\begin{tabular}{cc}
\begin{tabular}{@{}c@{}}
\includegraphics{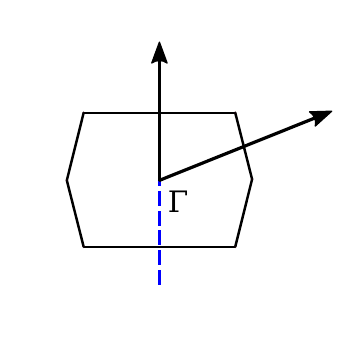} 
\end{tabular} &
\begin{tabular}{@{}c@{}}
\textbf{cm} ($\mathbf{D_1}$) \\ (Rhombohedral) \\\\
$\Gamma$: All\\
\end{tabular} 
\end{tabular}
\\
\hline
\begin{tabular}{@{}c@{}}
\includegraphics{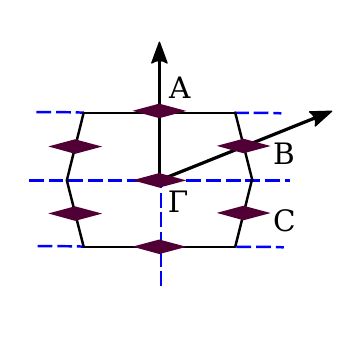} 
\end{tabular} &
\begin{tabular}{@{}c@{}}
\textbf{cmm} ($\mathbf{D_2}$), \textbf{cm*} \\ (Rhombohedral) \\\\
$\Gamma$: $A_{2n-1}$\\
A: $A_{2n-1}$\\
B: $A_{2n-1}$\\
C: $A_{2n-1}$
\end{tabular} 
&
\begin{tabular}{@{}c@{}}
\includegraphics{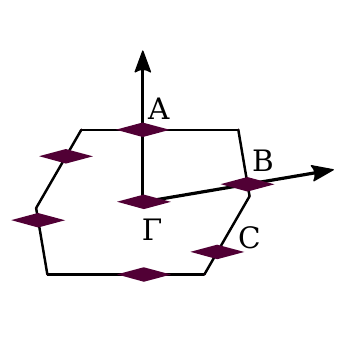} 
\end{tabular} &
\begin{tabular}{@{}c@{}}
\textbf{p2} ($\mathbf{C_2}$), \textbf{p1*} \\ (Oblique) \\\\
$\Gamma$: $A_{2n-1}$\\
A: $A_{2n-1}$\\
B: $A_{2n-1}$\\
C: $A_{2n-1}$
\end{tabular}
&
\begin{tabular}{@{}c@{}}
\includegraphics{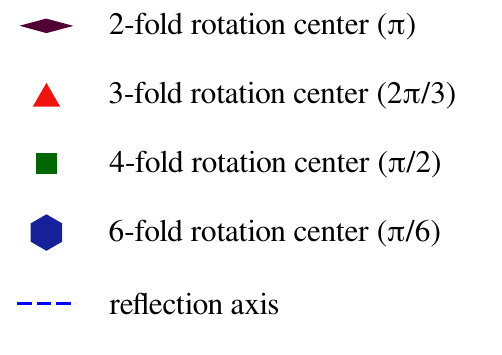} 
\end{tabular}
\\
\hline
\end{tabular}
\end{center}
\caption{In two dimensions, there are seventeen wallpaper groups which serve as symmetry groups of crystals. As a consequence of Bloch's theorem, the point group of the wallpaper group (mentioned above within parenthesis) acts about the origin in $k$-space. By combining the point group elements with the reciprocal lattice translations, we find the high symmetry points in the Brillouin zone and list the catastrophes  that are allowed at each of these points. If the Hamiltonian has time reversal invariance and no spin orbit coupling, there is an additional inversion symmetry in $k$-space. Such cases are denoted above with a * (For example \textbf{p1*}).}
\label{table4}
\end{table*}

\section{Identifying a singularity in practice}\label{sec:method}

In practice, the singularities need not occur in their standard form. Each of the 17 singularities forms a distinct equivalence class. (Equivalence here is up to a smooth, \emph{local} coordinate transformation with a smooth inverse). So we generally expect some smooth equivalent of the standard forms to occur in the context of band dispersions. But it might be hard to identify the singularity from such a non-standard expression. While we could try exhibiting a smooth coordinate transformation that puts the singularity in its standard form, it is very hard to do that for a generic dispersion. This apparent complication is easily resolved by exploiting the fact that codimension, determinacy, corank and winding are invariant under smooth coordinate transformations. By computing these numbers for the given singular dispersion we unambiguously identify its type. Below we give an explicit method for computing determinacy and codimension.

\subsection{Algorithm for computing determinacy}\label{sec:algorithm}
We now describe the method for computing determinacy and codimension and illustrate it with an example. The example has been chosen to illustrate the subtle aspects of Taylor expansions and how truncating the Taylor series to an incorrect order can yield wrong results which might appear sensible. We start by giving a few definitions and notations. The $k^{\mathrm{th}}$ degree Taylor polynomial of $f$ denoted by $j^k [f]$ or simply $j^k f$, is the truncation of the Taylor expansion of $f$ at the origin to degree $k$. A function $f$ is said to be $k$-\emph{determinate} if any function having the same $k^{\mathrm{th}}$ degree Taylor polynomial as $f$ is smoothly equivalent to it. In particular of course, $j^k f$ is also smoothly equivalent to $f$. Determinacy is then the lowest $k$ for which $f$ is $k$-determinate. \\

The method given below is a simplification of the method given in~\citep{castrigiano2004catastrophe}. It is also similar to the methods given in~\citep{poston2014catastrophe}. (See Appendix~\ref{app:method_explanation} for an explanation of the method).\\

\begin{enumerate}
\item Choose a particular $k$ to start. Determine the $k^{\mathrm{th}}$ degree Taylor polynomial $j^kf$.
\item Compute all polynomials of the form $k_x^m k_y^n \frac{\partial j^k f}{\partial k_x}$ and $k_x^m k_y^n \frac{\partial j^k f}{\partial k_y}$ with $1 \leqslant m+n \leqslant k$ and truncate them to degree $k$. There is only a finite number of such polynomials and we denote them by $p_1,\cdots,p_N$. These can be thought of as vectors in the ploynimial space spanned by the set $\{k_x^m k_y^n; 1 \leqslant m+n \leqslant k \}$.
\item Check if the system of linear equations $\sum_{i = 1}^{N} c_{i} p_i = k_x^j k_y^{k-j}$ has a solution for each $j = 0,\cdots,k$. 
\item If a solution exists for each of the systems, $f$ is $k$-determinate. We keep reducing $k$ by $1$, repeating the algorithm for each $k$ until we find the smallest $k$ for which the systems of equations all have a solution. Call it $k_{\mathrm{low}}$ and terminate the algorithm. 
\item If solution does not exist for each of the systems, we keep increasing $k$ by $1$, applying the algorithm for each $k$ until we find the first $k$ for which all of the systems of equations have a solution. Call it $k_{\mathrm{low}}$ and terminate the algorithm.
\end{enumerate}
Once the algorithm terminates successfully and $k_{\mathrm{low}}$ is determined, it can be shown that the determinacy of $f$ is either $k_{\mathrm{low}}$ or $k_{\mathrm{low}}-1$. While this ambiguity cannot be resolved easily, it is not a problem as long as we determine the other indices unambinguously. This is because no two of the seventeen singularities whose determinacy differs by $1$ have the same corank, codimension and winding.\\ 

Let us now give a method for finding the codimension. For $k = k_{\mathrm{low}}$, construct a matrix $M$ by listing each of $p_1,\cdots,p_N, \frac{\partial j^k f}{\partial k_x}, \frac{\partial j^k f}{\partial k_y}$ as row vectors in the basis of the monomials $\{k_x^m k_y^n; 1 \leqslant m+n \leqslant k \}$. Let $\mathrm{rank}(M)$ denote the rank of $M$. The codimension is then $(k^2+3k)/2-\mathrm{rank}(M)$.

\subsection{A sample computation}
\label{sec:determinacy}
\begin{figure*}[!ht]
 \centering
 \includegraphics[width=1\textwidth]{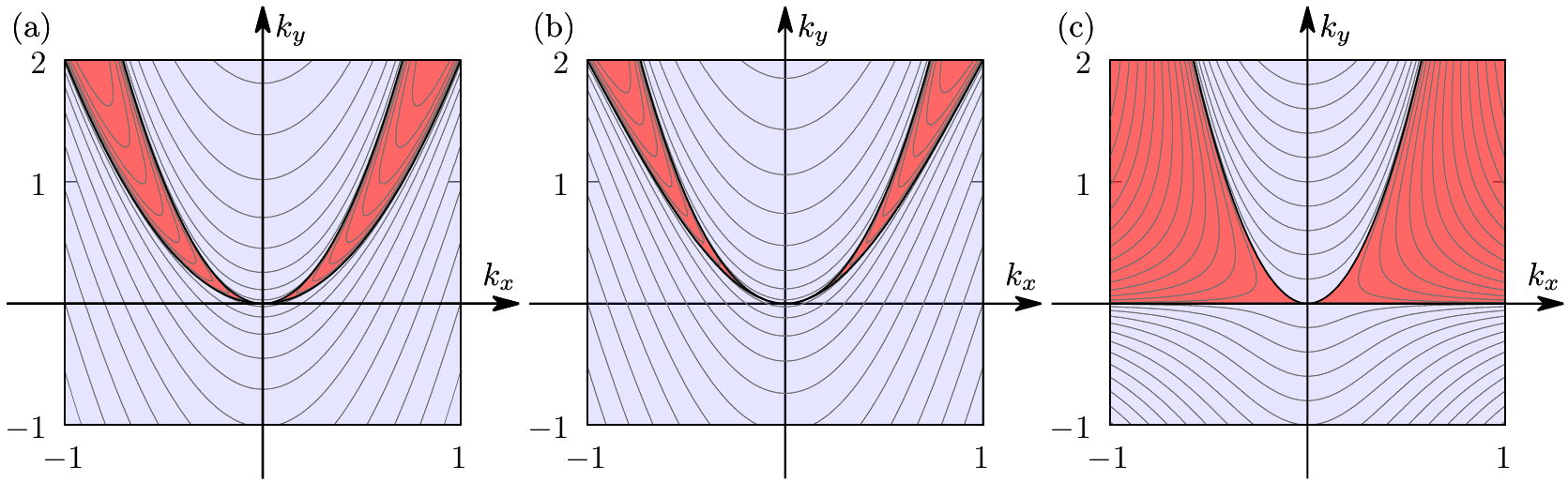}
 \caption{Constant energy contours of the dispersions: (a) $f(k_x,k_y) = -k_y^2+6 k_x^2 k_y - 8 k_x^4$, (b) $g(k_x, k_y) = -k_y^2 + 6 k_x^2 k_y -9 k_x^4 + k_x^6$ and (c) $h(k_x, k_y) = -k_y^2 + 6 k_x^2 k_y$ shown within the range $-1 < k_x < 1$, $-1 < k_y < 2$. While the energy contours of $f$ and $g$ appear to have the same topology, they are actually not related by smooth deformation. The former corresponds to two parabolas that intersect tangentially and the latter consists of two cubic curves that intersect tangentially at $(0,0)$. The DOS diverges differently for both these dispersions although they give the same cubic truncation $h$. This in spite of the fact that $h$ appears to produce correct topology of Fermi surface (two curves tangentially touching at an \emph{isolated} higher order critical point) and a seemingly sensible power law divergence of DOS. This is shown to be the consequence of the fact that $f$ and $g$ are inequivalent and have different allowed order of truncation as indicated by their determinacy.}
 \label{fig:example2}
\end{figure*}

Consider two functions $f(k_x,k_y) = -k_y^2+6 k_x^2 k_y - 8 k_x^4$ and $g(k_x,k_y) = -k_y^2 + 6 k_x^2 k_y -9 k_x^4 + k_x^6$. On cubic truncation, both these functions yield $h(x) = -k_y^2 + 6 k_x^2 k_y$. This function arises for example in the context of the continuum model for twisted bilayer graphene in~\cite{1901.05432}. The Fermi surfaces of the three functions are depicted in Fig~\ref{fig:example2}. Visually, they appear to share the same topology for the energy contours. In particular, the zero energy contour seems to consist of two curves intersecting tangentially at the origin. Naively this might motivate truncation to cubic order for computing the DOS among other things. We now demonstrate that this would lead to an incorrect result, at least for $g$, by showing that while $f$ and $h$ are equivalent to the cusp, $g$ is equivalent to butterfly. \\

We first compute the determinacy of $f$. Let $\overline{q}^{\, n}$ denote the truncation of the polynomial $q$ to degree $n$. Choosing $k=4$ we list some of the polynomials $p_i$ for this function:
\begin{align*}
& p_1 = \overline{k_x \frac{\partial j^4 f}{\partial k_x}}^{\, 4} = 12 k_x^2 k_y - 32 k_x^4\\
& p_2 = \overline{k_x \frac{\partial j^4 f}{\partial k_y}}^{\, 4} = -2 k_x k_y + 6 k_x^3 \\
& p_3 = \overline{k_y \frac{\partial j^4 f}{\partial k_x}}^{\, 4} = 12 k_x k_y^2 - 32 k_x^3 k_y \\
& p_4 = \overline{k_y \frac{\partial j^4 f}{\partial k_y}}^{\, 4} = -2 k_y^2 + 6 k_x^2 k_y \\
& p_5 = \overline{k_x^2 \frac{\partial j^4 f}{\partial k_x}}^{\, 4} = 12 k_x^3 k_y \\
& p_6 = \overline{k_x k_y \frac{\partial j^4 f}{\partial k_x}}^{\, 4} = 12 k_x^2 k_y^2 \\
& p_7 = \overline{k_y^2 \frac{\partial j^4 f}{\partial k_x}}^{\, 4} = 12 k_x k_y^3 \\
& p_8 = \overline{k_x^2 \frac{\partial j^4 f}{\partial k_y}}^{\, 4} = -2k_x^2 k_y + 6 k_x^4 \\
& p_9 = \overline{k_y^3 \frac{\partial j^4 f}{\partial k_y}}^{\, 4} = -2 k_y^4
\end{align*}
The common factors will not affect the results and we have dropped them. It immediately follows that 
\begin{eqnarray}
\frac{1}{4} \left( p_1 + 6 p_8 \right)  = k_x^4 \\
\frac{1}{12} p_5 = k_x^3 k_y \\
\frac{1}{12} p_6 = k_x^2 k_y^2 \\
\frac{1}{12} p_7 = k_x k_y^3 \\
-\frac{1}{2} p_8 = k_y^4 \\
\end{eqnarray}
Thus $f$ is $4$ determinate. When we generate $p_i$ for $k=3$, we find that no linear combination $\sum_i c_i p_i$ will yield $k_x^3$. Thus $k = 4$ is the smallest $k$ for which the algorithm works implying $\det [f] = 4$ or $\det [f] = 3$. It can easily be shown that $\mathrm{cor} [f] = 1$, $\mathrm{w} [f] = 4$ and $\mathrm{cod} [f] = 2$. This verifies that $f$ is equivalent to cusp. Proceeding in the same way we can show that $\det [g] = 6$ rather than $4$ and that it is equivalent to butterfly. This clearly shows that $f$ and $g$ are \emph{different} singularities inspite of the similarity of their Fermi contours. \\ 

Finally it can be shown that $h$ is also equivalent to the cusp. This naturally leads to the question as to why $g$ is also not equivalent to the cusp given that its cubic truncation is $h$, which is equivalent to the cusp. The answer lies in the definition of determinacy. The relation $\det [h] = 4$ only guarantees that functions whose fourth degree Taylor polynomials match that of $h$ are equivalent to it and therefore the cusp. This importantly requires such functions to have no quartic terms in their Taylor expansion as $h$ does not. Since $g$ fails to satisfy this property, it is not guaranteed to be equivalent to $h$. This also means that from the point of view of the mathematical theory, the truncation of $f$ to cubic order is also not justified although in this specific instance it worked.


\section{Summary and outlook}\label{sec:summary}

In this paper we presented a classification of possible Fermi surface
topological transitions in two spatial dimensions that take place via
higher order singularities. At these higher order singularities, the
DOS diverges as a power law, in contrast to the weaker
logarithmic divergence at van Hove singularities. When the number of
control parameters is less than or equal to seven, we can list all the seventeen
possibilities classified by catastrophe theory. The critical exponents
controlling the divergence of the DOS, as well as the
ratio of the prefactor or amplitudes for energies on both sides of the
transition, are universal properties associated to the specific type
of catastrophe. Smooth coordinate transformations cannot change these
exponents or ratios, as well as the topology of the critical Fermi
surface right at the singularity.

We further showed that by tuning the parameters in the Hamiltonian, we can obtain higher order singularities. They are especially easy to achieve at high symmetry points in the Brillouin zone. At the same time, the high symmetry points also restrict the types of singularities: only those consistent with the symmetry can occur. Motivated by this, we classified the singularities that are likely to occur at high symmetry points in the Brillouin zones corresponding to the seventeen wallpaper groups in two dimension.	

These critical points of the electronic dispersion can serve as the
starting point for the treatment of interactions, and can potentially
lead to novel electronic phenomena. Besides simply facilitating
ordering transitions because of the enhancement of the DOS, there are several more exotic possibilities that demand
exploration. While the addition of weak interactions on smooth Fermi
surfaces leads to a Landau Fermi liquid, it is not clear that
singularities in the curvature of the Fermi surface and their
associated power law enhancement of the electronic DOS
may not lead to novel quantum critical behavior. In the singularities
that we discussed, the singular point is not isolated (i.e., a
point-like Fermi surface); there is a finite Fermi surface containing
this point. So the analysis of interactions for certain singularities
may require dealing with singular and non-singular parts of the Fermi
surface. For instance, the singular points with their divergent
DOS may enhance scattering and reduce quasi-particle
lifetime; at the same time, the Fermi velocities vanish at those
points. Therefore, properties like transport, which depend both on the
Fermi velocities and quasiparticle lifetimes, may depend on the joint
effects of the singular and non-singular parts of the Fermi
surface. Understanding the effects of interactions on the higher order
singularities is an interesting and largely open problem; while some
progress has been recently made~\citep{PhysRevB.95.035137, 1810.13392,
1901.05432}, much remains to be understood.

In addition to the understanding of the classes of Fermi surface
topological transitions, it should be possible to apply the machinery
of singularity and catastrophe theory to other aspects of electronic
dispersions. Here we focused on point singularities, but it may be
fruitful to develop a catalog of line singularities as well. For
example, in tight-binding Hamiltonians for the diamond lattice, there
are cases where the Fermi surfaces collapse to lines. A step towards
these generalizations would be to start with polynomials in $k_x,k_y$
that multiply Pauli matrices, instead of just scalars. The matrix
structure should bring about square root singularities that are not
present in the simple scalar dispersions. We believe that the study
here presented is one step in understanding singular behavior
in electronic systems by deploying the tools of catastrophe theory.


\textbf{Note added:} While we were completing this manuscript, we
became aware of similar work also being completed by Noah F. Q. Yuan
and Liang Fu. We point the reader to that reference as well. We thank
Zhi-Cheng Yang for recognizing that the works were similar, and
bringing the two groups into contact.

\acknowledgements{
This work is supported by the DOE Grant
No. DE-FG02-06ER46316 (A.C. and C.C.), the EPSRC grant No. EP/P002811/1 (J.J.B.) and the Royal Society  (J. J. B. and C.C.). We thank Siddhant Das for useful discussions on integrals.
}

\appendix


\section{Density of States}
\label{app:DOS}
\subsection{Invariance under smooth transformations}

The DOS in the $n^{\mathrm{th}}$ band ignoring spin degeneracy is defined as~\cite{Ashcroft}
\begin{equation}
g(\epsilon) = \int_{\mathrm{BZ}} \frac{d^2 \tilde{\bm{k}}}{(2\pi)^2} \,\, \delta(\tilde{E}_n(\tilde{\bm{k}}) - \epsilon) = \int_{S_n(\epsilon)} \frac{dl}{(2\pi)^2} \, \frac{1}{\| \nabla \tilde{E}_n(\tilde{\bm{k}}) \|}
\end{equation}
Where $S_n(\epsilon)$ is the constant energy contour in the $n^{\mathrm{th}}$ band. If for a particular energy, the constant energy contour contains a critical point, the integrand diverges at that point.

Let us drop $n$ so that the given dispersion is $\tilde{E}(\tilde{\bm{k}})$. Assume that under a suitable \emph{local} smmothly reversible coordinate transformation $\tilde{\bm{k}} \rightarrow \bm{k}$, the dispersion transforms to $E(\bm{k})$, one of the standard types. The Jacobian determinant for the coordinate transformation can be Taylor expanded as:

\begin{equation}
J(k_x,k_y) = J_0 + \sum_{n=1}^{\infty} \left(\sum_{m=1}^n c_{nm} \, k_x^m k_y^{n-m} \right) ,
\end{equation} 
where $J_0 \neq 0$ since that condition is equivalent to the coordinate transformation being a local diffeomorphism. Since the required coordinate transformation will in general be local, we restrict the DOS integral to a bounded neighborhood $D$ of the origin. This procedure will not affect the calculation of the divergent part of the integral since it arises due to the critical point at the origin. (In fact we can also extend the integral to the entire two-dimensional plane). The DOS now reads:
\begin{multline}
g(\epsilon)=\int_{D} \frac{d^2 k}{(2\pi)^2} J(k_x,k_y) \, \delta(E(k_x,k_y) - \epsilon) \\
= J_0 \int_{D} \frac{d^2 k}{(2\pi)^2}  \delta(E(k_x,k_y) - \epsilon) \\
+ \sum_{n=1}^{\infty} \left(\sum_{m=1}^n c_{mn}  \int_{D} \frac{d^2 k}{(2\pi)^2} k_x^m k_y^{n-m} \, \delta(E(k_x,k_y) - \epsilon) \right) 
\label{eq:DOSseries}
\end{multline}
For dispersions which are either homogeneous or take the form $E(\bm{k}) = a \, k_x^{m_1} k_y^{n_1} + b \, k_x^{m_2} k_y^{n_2}$ for non-negative integers $m_i$ and $n_i$, it might be possible to scale $k_x \rightarrow |\epsilon|^{\alpha} k_x$ and $k_y \rightarrow |\epsilon|^{\beta} k_y$ for appropriate positive rational numbers $\alpha$ and $\beta$ so that $d^2k \, \delta (E(k_x,k_y)-\epsilon) \rightarrow |\epsilon|^{-\gamma} [d^2k \, \delta (E(k_x,k_y)\pm 1)]$, ($-1$ if $\epsilon > 0$, $+1$ is $\epsilon < 0$). For instance, for a homogeneous dispersion of degree $n$, $\alpha = \beta = 1/n$ while for $E(\bm{k}) = a \, k_x^{m_1} k_y^{n_1} + b \, k_x^{m_2} k_y^{n_2}$, we set $\alpha = (n_2 - n_1)/(m_1 n_2 - m_2 n_1)$ and $\beta = (m_1 - m_2)/(m_1 n_2 - m_2 n_1)$. (For the standard types which take the latter form, $m_1 n_2 - m_2 n_1 \neq 0$ so that this scaling is allowed). Performing this scaling in Eq.~\ref{eq:DOSseries}, we get:
\begin{multline}
g(\epsilon)= |\epsilon|^{-\gamma} \bigg[ J_0 \int_{D} \frac{d^2 k}{(2\pi)^2}  \delta(E(k_x,k_y) \pm 1) \\
+ \sum_{n=1}^{\infty} \sum_{m=1}^n \bigg( c_{mn} |\epsilon|^{m\alpha + (n-m)\beta} \\ \times \int_{D} \frac{d^2 k}{(2\pi)^2} k_x^m k_y^{n-m} \, \delta(E(k_x,k_y) - \pm 1) \bigg) \bigg]
\label{eq:DOSseries2}
\end{multline}
Since $\alpha, \beta, \gamma > 0$ and $n \geqslant m$, we have $m\alpha + (n-m)\beta > 0$ so that in the limit $\epsilon \rightarrow 0$ the leading order divergent part of the integral is given by:
\begin{equation}
g(\epsilon) \sim \bigg(J_0 \int_{\mathbf{R}^2} \frac{d^2 k}{(2\pi)^2}  \delta(E(k_x,k_y) \pm 1) \bigg) |\epsilon|^{-\gamma}
\end{equation}
where once again we choose $-1$ if $\epsilon > 0$, $+1$ is $\epsilon < 0$. We have extended the integral to the entire plane since this only adds only a finite error that does not change the coefficient of the leading order divergent part. Thus, we infer that the exponent $\gamma$ and the ratio $D_+/D_-$ are invariant under a smoothly reversible coordinate transformation. We compute $D_+/D_-$ for the various catastrophes in the forthcoming sections.
\subsection{Cuspoid catastrophes}
The cuspoid catastrophes include the fold, cusp, swallowtail, butterfly, wigwam and star whose dispersions take the form: $\epsilon(\bm{k}) = k_x^n - k_y^2$. For the sake of notational simplicity, in what follows we replace $k_x \rightarrow x$ and $k_y \rightarrow y$. We now derive the DOS for these. \\\\
Consider $\epsilon (x,y) = x^n - y^2$. This gives $y = \pm \sqrt{x^n - \epsilon}$. Furthermore:
\begin{multline}
\frac{dy}{dx} = \pm \frac{n}{2} \frac{x^{n-1}}{\sqrt{x^n - \epsilon}} \implies dl = dx \sqrt{1+ \bigg( \frac{dy}{dx} \bigg)^2} \\ = dx \, \frac{1}{2} \frac{\sqrt{n^2 x^{2n-2} +4 x^n -4 \epsilon}}{\sqrt{x^n - \epsilon}}
\end{multline}
Now $\nabla \epsilon = (n x^{n-1}, -2y)$. This gives $\| \nabla \epsilon \| = \sqrt{n^2 x^{2n-2} +4y^2} = \sqrt{n^2 x^{2n-2} +4x^n -4\epsilon}$, where we have used the dispersion to substitute for $y^2$. 

Now let $n$ be even. For $\epsilon < 0$, the constant energy curves have vertices $(0,\pm \sqrt{|\epsilon |})$ disperse upward/downward in the $xy$-plane. The DOS integral then becomes:
\begin{equation}
g(\epsilon) = \int_{S(\epsilon)} \frac{dl}{(2\pi)^2} \, \frac{1}{\| \nabla \epsilon_n(\bm{k}) \|} = \frac{1}{4\pi^2} \int_{-\infty}^{\infty} \frac{dx}{\sqrt{x^n - \epsilon}}
\end{equation}
Where we have substituted for $dl$ and $\| \nabla \epsilon \|$ in the integrand and simplified the expression. We have also accounted for the fact that the upward and downward dispersing branches contribute equally to the integral. Now we write $-\epsilon = \delta^n$, with $\delta > 0$, (so that $\delta = |\epsilon|^{\frac{1}{n}}$). We then make the substitution $x = \delta y$. These then give:
\begin{multline}
g(\epsilon) = \frac{1}{4\pi^2} \int_{-\infty}^{\infty} \frac{\delta \, dy}{\delta^{\frac{n}{2}} \sqrt{y^n + 1}} \\= \bigg(\frac{1}{2\pi^2} \int_{0}^{\infty} \frac{dy}{\sqrt{y^n + 1}} \bigg) |\epsilon|^{\frac{1}{n} - \frac{1}{2}} , \,\,(\epsilon < 0)
\end{multline}
For $\epsilon > 0$, the constant energy curves disperse rightward/leftward. The vertices are at $(\pm \epsilon^{\frac{1}{n}}, 0)$. The DOS integral in this case is:
\begin{multline}
g(\epsilon) = \frac{1}{2\pi^2} \int_{\epsilon^{\frac{1}{n}}}^{\infty} \frac{dx}{\sqrt{x^n - \epsilon}} \\= \bigg( \frac{1}{2\pi^2} \int_1^{\infty} \frac{dy}{\sqrt{y^n - 1}} \bigg) |\epsilon|^{\frac{1}{n} - \frac{1}{2}} , \,\,(\epsilon > 0)
\end{multline}
Where once again we have used the same set of substitutions to simplify the integrand. There are four four branches of the contour which contribute equally to the above integral. For $\epsilon = y^2 - x^n$ with $n$ even, the roles of positive and negative $\epsilon$ are interchanged in the above derivation. \\

For odd $n$ the constant energy contours of $\epsilon = x^n - y^2$ disperse rightward for both $\epsilon \leqslant 0$ and $\epsilon > 0$. When $\epsilon < 0$, the vertex of the contour is $(-|\epsilon|^{\frac{1}{n}},0)$. The DOS is given by:
\begin{multline}
g(\epsilon) = \frac{1}{4\pi^2} \int_{-|\epsilon|^{\frac{1}{n}}}^{\infty} \frac{dx}{\sqrt{x^n-\epsilon}} \\ = \bigg( \frac{1}{4\pi^2}\int_{-1}^{\infty} \frac{dy}{\sqrt{y^n + 1}} \bigg) |\epsilon|^{\frac{1}{n} - \frac{1}{2}}, \,\, (\epsilon < 0)
\end{multline}
Where once again we have scaled $x$ to extract out the power law dependence. On the contrary, when $\epsilon > 0$, the vertex is at $(|\epsilon|^{\frac{1}{n}},0)$ and the DOS is given by:
\begin{multline}
g(\epsilon) = \frac{1}{4\pi^2} \int_{|\epsilon|^{\frac{1}{n}}}^{\infty} \frac{dx}{\sqrt{x^n-\epsilon}} \\= \bigg(\frac{1}{4\pi^2} \int_{1}^{\infty} \frac{dy}{\sqrt{y^n - 1}} \bigg) |\epsilon|^{\frac{1}{n} - \frac{1}{2}}, \,\, (\epsilon > 0)
\end{multline}
The dispersion $\epsilon = y^2 - x^n$ with $n$ odd just has the role of positive and negative energies interchanged in the above. In fact $\epsilon = x^n + y^2$ is also equivalent to the above case. This is because, the coordinate transformation $x \rightarrow -x$ does not affect the DOS. \\

Thus, $g(\epsilon) \sim |\epsilon|^{\frac{1}{n} - \frac{1}{2}}$ for the cuspoid catastrophes although the coefficient in front depends on the dispersion.

\subsection{The umbilics and the rest}
We first treat the hyperbolic umbilic and elliptic umbilic catastrophes and their higher order generalizations. These respectively take the form $\epsilon(x,y)=x^2 y \pm y^n$ for odd $n$. The energy contours of these are symmetric in the sense that the transformation $y \rightarrow -y$ takes us from the contour of $\epsilon$ to the contour of $-\epsilon$. From this property, it is easy to see that the coefficients $D_+$ and $D_-$ are equal so that $D_+/D_- = 1$. Since we are ultimately interested only in their ratio, we will not bother evaluating them and just apply the appropriate scaling procedure to extract the exponent. To this end consider $\epsilon > 0$ and set $x = \epsilon^{\frac{1}{2}-\frac{1}{2n}}$ and $y = \epsilon^{1/n} v$. The Jacobian determinant for this transformation is:
\begin{equation}
\det \frac{\partial (x,y)}{\partial (u,v)} = \epsilon^{\frac{1}{2}+\frac{1}{2n}}
\end{equation}
Using the scaling property of the delta function, viz $\delta (ax) = \delta(x) / |a|$ the DOS integral becomes:
\begin{equation}
g(\epsilon) = \int du \,dv \, \epsilon^{\frac{1}{2}+\frac{1}{2n}} \, \frac{\delta(u^2 v \pm v^n - 1)}{\epsilon} \propto |\epsilon|^{\frac{1}{2}-\frac{1}{2n}}
\end{equation}
The parabolic umbilic and its generalizations take the form $x^2 y + y^n$ for even $n$. We do not separately consider the case $x^2y-y^n$ since it can be obtained from the former by the transformation $y \rightarrow -y$ followed by an overall sign change. We proceed just as in the case of the cuspoids to compute $D_+/D_-$.


\section{Symmetries and critical points}
\label{app:symmetry}
We now compute the effect of reflection (mirror) symmetry and rotational symmetry on the Jacobian. First consider reflection about the $k_x$ axis. This is represnted by:
\begin{equation}
r_{k_x} = \begin{pmatrix}
1 & 0 \\[5pt]
0 & -1
\end{pmatrix}
\end{equation}
Now for a coordinate transformation $\phi$ the Jacobian at the origin transforms as $Df(\phi(0)) = Df(0) D\phi(0)$. Let $\bm{k} = (k_x, k_y)$. Since $f(\bm{k}) = f(r_{k_x} \cdot \bm{k})$, we have:
\begin{equation}
\begin{pmatrix}
\frac{\partial f}{\partial k_x} &
\frac{\partial f}{\partial k_y}
\end{pmatrix}
=
\begin{pmatrix}
\frac{\partial f}{\partial k_x} &
\frac{\partial f}{\partial k_y}
\end{pmatrix}
\begin{pmatrix}
1 & 0 \\[5pt]
0 & -1
\end{pmatrix}
=
\begin{pmatrix}
\frac{\partial f}{\partial k_x} &
-\frac{\partial f}{\partial k_y}
\end{pmatrix}
\end{equation}
This implies $\partial f/ \partial k_y = 0$ at the origin. Thus, the direction perpendicular to the mirror has a vanishing component for the Jacobian. Now let $\rho_{\theta}$ denote rotaion by $\theta$ in the plane. Applying the same procedure as above gives:
\begin{multline}
\begin{pmatrix}
\frac{\partial f}{\partial k_x} &
\frac{\partial f}{\partial k_y}
\end{pmatrix}
=
\begin{pmatrix}
\frac{\partial f}{\partial k_x} &
\frac{\partial f}{\partial k_y}
\end{pmatrix}
\begin{pmatrix}
\cos \, \theta & -\sin \, \theta \\[5pt]
\sin \, \theta & \cos \, \theta
\end{pmatrix} \\
\implies \begin{pmatrix}
(\cos \, \theta - 1) & \sin \, \theta \\[5pt]
-\sin \, \theta & (\cos \, \theta - 1)
\end{pmatrix}
\begin{pmatrix}
\frac{\partial f}{\partial k_x} \\[5pt]
\frac{\partial f}{\partial k_y}
\end{pmatrix} = 0
\end{multline}
The determinant of this homogeneous system is $4 \sin^2 \, \frac{\theta}{2}$, which is non zero for $\theta = \pi/3, \pi/2, \pi, 2\pi/3$ which are generators of non-trivial lattice rotation symmetries. This implies the unique solution is $\partial f/\partial k_x = \partial f/\partial k_y = 0$, so that we have a critical point.


\section{Quick review of catastrophe theory}\label{app:catastrophe_review}

The discussion in this section closely follows ~\cite{poston2014catastrophe, castrigiano2004catastrophe}. We first review a few definitions and conventions. If $U$ is an open subset of $\mathbb{R}^n$, a function $f:U \rightarrow \mathbb{R}$ is smooth if derivatives of all orders exist. From here on $f$ will always denote a smooth function defined on $U$, an open subset of $\mathbb{R}^n$. We shall use map and coordinate transform interchangeably to refer to a bijective function $\phi : U \rightarrow V$, where $U$ and $V$ are open subsets of $\mathbb{R}^n$.\\ 
The derivative of $f$ at any point $x_0 \in U$, denoted by $Df|_{x_0} : \mathbb{R}^n \rightarrow \mathbb{R}$, is a linear transformation. When it acts on a vector, it gives the derivative of the function with respect to the vector; in particular, unit vectors give directional derivatives. Also, the function $f(x_0) + Df|_{x_0} \cdot (x-x_0)$ is the best linear approximation to $f$, near $x_0$ (i.e., in some neighborhood of $x_0$). These notions extend to higher derivatives as well. We note that if $x$ is represented by a column vector, then $Df|_{x_0}$ is represented by a row vector (that we refer to as the Jacobian), while the second derivative, denoted by $D^2 f|_{x_0}$ is an $n \times n$ matrix called the Hessian.\\
$x_0$ is a critical point if $Df|_{x_0}$ vanishes. Since, from now on we shall focus on critical points, we assume that the origin has been translated so that the critical point we are interested in occurs at the origin. 

\subsection{Diffeomorphism}
Let $V \subset \mathbb{R}^n$ be open. A bijective map $\phi: U \rightarrow V$ is a diffeomorphism if both $\phi$ and $\phi^{-1}$ are smooth. A smooth map $\psi : U \rightarrow \mathbb{R}^n$ is a \emph{local} diffeomorphism at $x_0 \in U$ if there exists an open neighborhood $V$ of $x_0$ such that $\psi$ restricted to $V$ is a diffeomorphism. It can be shown that $\phi$ is a local diffeomorphism at $x_0$ if and only if $\det D\phi|_{x_0} \neq 0$.

\subsection{Equivalence of functions}
Two funtions $f$ and $g$ are \emph{equivalent} if there exists a diffeomorphism $\phi: U \rightarrow V$ and a constant $\gamma$ (called the shear term) such that $ g(x) = f(\phi(x)) + \gamma$, $\forall x \in U$. We shall often refer to such a $\phi$ as a smoothly reversible coordinate transformation.

\subsection{Morse and non-Morse critical points - corank}
\label{sec:corank}
If the Hessian of $f$ at origin is nondegenerate (i.e., it's rank equals $n$), then Morse lemma tells us that in some neighborhood of the origin, $f$ is equivalent (in the above sense) to a quadratic form $-y_1^2- \cdots -y_l^2+y_{l+1}^2+ \cdots +y_n^2$. A quadratic form having this structure is referred to as a \emph{Morse l-saddle}. (Thus, a maximum is an n-saddle while a minimum is a 0-saddle). The critical point is itself referred to as a Morse critical point or an \emph{ordinary critical point}. Another way to state Morse lemma is to say that near an ordinary critical point, there is a local coordinate system in which the function takes the form of a Morse l-saddle.\\

When the Hessian is degenerate and has rank $r$ (and therefore corank $n-r$), then we can apply the splitting lemma which states that the function is equivalent near $0$ to a function of form $\pm x_1^2 \pm \cdots \pm x_r^2 + \hat{f} (x_{r+1}, ..., x_n)$. We refer to $\pm x_1^2 \pm \cdots \pm x_r^2 $ as the Morse part and $\hat{f} (x_{r+1}, ..., x_n)$ as the non-Morse part or residual singularity. The critical point is itself referred to as a non-Morse critical point or a higher order critical point. The corank of the function, denoted by $\mathrm{cor} [f]$ is the number of zero eigenvalues of the Hessian.

\subsection{Structural stability}
We say f is structurally stable if for sufficiently small perturbations $p$, $f$ is equivalent to $f+p$. We know that non-degeneracy of the Hessian is equivalent to the condition $\det D^2f|_0 \neq 0$. Since determinant is a continuous function, we expect that when we add a small perturbation $p$, $\det D^2(f+p)|_0$ is still non-zero in some neighborhood of 0, so that the critical point (which might have moved), is still Morse. In fact the converse is also true so that we have the following result: A critical point is structurally stable if and only if it is Morse.

\subsection{Jets and jet spaces}
The $k$-jet of a smooth function, denoted by $j^k f$, is obtained by taking the Taylor series up to degree $k$ (i.e., truncating terms of $O(k+1)$ and higher). The vector space $J_n^k$ is the set of all polynomials of degree $k$ in $n$ variables, with \emph{zero} constant term. (alternatively $J_n^k = \{j^k f \, | \, \mathrm{all} \,\, f: \mathbb{R}^n \rightarrow \mathbb{R} \,\, \mathrm{with} \,\, f(0) = 0 \}$).

\subsection{Determinacy}
We say that a two functions $f$ and $g$ are $k$-equivalent if $j^k f = j^k g$. A function $f$ is $k$-determined if every function $k$-equivalent to it is also equivalent to it. In particular we have $j^k[j^k f] = j^k f$ so that $j^k f \sim f$. This means that if $f$ is $k$-determined, we can find a smoothly reversible coordinate transformation that maps $f$ to $j^k f$, effectively removing terms of $O(k+1)$ and higher from it's Taylor expansion.

The determinacy of a function, denoted by $\det [f]$ is the smallest $k$ for which $f$ is $k$-determinate. If there is not finite such $k$ then $\det [f] = \infty$. The determinacy of a function is preserved under a smoothly reversible coordinate transformation.

\subsection{Codimension}
We first define some polynomial spaces:\\
$E_n^k$ is the vector space of all polynomials in $x_1, \cdots , x_n$ of degree $\leqslant k$.\\
$J_n^k$ is the subspace of $E_n^k$ of polynomials with zero constant term.\\
$M_n^k$ is the subspace of all homogeneous polynomials of degree $k$.

The expression $\overline{P}^{\, k}$, where $P$ is a polynomial, refers to the truncation of $P$ to degree $k$.\\

The tangent space $\Delta_k(f)$ to $f$ is the subspace of $J_n^k$ spanned by polynomials  of the form $\overline{Q j^k \big( \frac{\partial f}{\partial x_i} \big),}^{\, k}$ where $Q \in E_n^k$. It contains the directions in which $j^k f$ moves under smooth coordinate transformations. It can be shown that $\Delta_k(f) = \Delta_k(j^{k+1}f)$.

The codimension of $f$, denoted by $\mathrm{cod}(f)$ is the codimension of $\Delta_k(f)$ in $J_n^k$, for \textit{any} $k$ for which $f$ is $k$-determinate. $\mathrm{cod}(f)$ tells us the number of missing directions in the subspace in which $j^k f$ moves. Thus, these directions, which are really polynomial terms, can not be added or removed by smooth coordinate changes. The codimension of a function is preserved under a smoothly reversible coordinate transformation.

\subsection{Winding number}
\label{sec:winding}
We introduce the notion of winding number which counts the number of times the function changes sign along a closed contour around the origin. Since this is an integer, we expect that it does not change under smooth coordinate transformation.

\subsection{Unfolding}
An $r$-unfolding of $f$ at $0$ is a function $F: U \times \mathbb{R}^{r} \rightarrow \mathbb{R}$ with $(x_1,\cdots,x_n, t_1,\cdots, t_r) \mapsto F(x,t) = F_t(x)$, such that $F_{0,\cdots,0}(x) = f(x)$. We also refer to it as an $r$-parameter family through $f$. In this context, the $x_i$ are state variables while $t_i$ are control variables.

A $d$-unfolding $\overline{F}$ is \emph{induced} from an $r$-unfolding $F$ via three mappings defined in a region around the origin so that $\overline{F}(x,s) = F(y_s(x),e(s)) + \gamma(s)$ where:\\
\begin{eqnarray*}
e:\mathbb{R}^d \rightarrow \mathbb{R}^r, \,\,\, (s_1,\cdots ,s_d) \mapsto (e_1 (s), \cdots , e_r (s)), \\
y:\mathbb{R}^{n+d} \rightarrow \mathbb{R}^n, (x,s) \mapsto (y_1 (x,s), \cdots ,y_n(x,s)), \\
\gamma : \mathbb{R}^d \rightarrow \mathbb{R}.
\end{eqnarray*}

An $r$-unfolding of $f$ at $0$ is versal if all other unfoldings of $f$ at $0$ can be induced from it. If further $r = \mathrm{cod}(f)$, then it is universal. Finally, we mention an important result: If $f$ is $k$-determinate, then we can construct a universal unfolding for $f$ by choosing a cobasis $\{ p_1(x),\cdots ,p_r(x) \}$ for $\Delta_k (f)$ in $J_n^k$ and defining:
\begin{equation}
F(x_1, \cdots ,x_n, t_1, \cdots, t_r) = f(x) + t_1 \, p_1(x) + \cdots + t_r \, p_r(x).
\end{equation}

\section{Thom's theorem basic catastrophes}
The original Thom's theorem states that any $r$-parameter family of smooth functions $\mathbb{R}^n \rightarrow \mathbb{R}$ for $r \leqslant 4$ and any $n \geqslant 1$ is typically structurally stable and equivalent to one of the following seven types:

\subsubsection{Fold}
The fold has a universal unfolding: $x^3 + t_1 x \pm y^2$. The determinacy is $3$ while codimension is $1$, so that the fold can be thought of as two ordinary critical points merging into a higher order critical points. Only symmetry present in the standard form is reflection about $y$-axis (i.e., $r_y$).

\subsubsection{Cusp}
Any universal unfolding of the cusp is equivalent to: $\pm(x^4 + t_1 x^2 + t_2 x) \pm y^2$. The determinacy is $4$ and since the codimension is $2$, under the action of the unfolding, three ordinary critical points merge to give a cusp. This could be preceeded by a situation wherein just two of the three critical points merge to give a fold and Morse critical point. This is prohibited if $t_2 = 0$, since in that case we have $r_x$ and $\rho_{\pi}$ symmetries in addition to $r_y$.

\subsubsection{Swallowtail}
The swallowtail has a determinacy of $5$ and is described by the unfolding: $x^5 + t_1 x^3 + t_2 x^2 + t_3 x \pm y$. With a codimension of $3$ it corresponds to the merging of $4$ ordinary critical points on a line. On the way to the swallowtail we can get cusps and folds. Similar to fold it has only $r_y$ as symmetry.

\subsubsection{Butterfly}
For the butterfly, the universal unfolding takes the form: $\pm (x^6 + t_1 x^4 + t_2 x^3 + t_3 t^2 + t_4 x) \pm y^2$. This has the consequence that the $5$ ordinary critical points lying on the $x$-axis merge to give higher order critical point(s). The unfolding can also exhibit the fold, cusp and swallowtail due to the merging of fewer than five critical points. When $t_2 = 0$ and $t_4 = 0$, it has $r_x$ and $\rho_{\pi}$ as symmetries in addition to $r_y$.

\subsubsection{Elliptic umbilic}
The standard form of the unfolding for the elliptic umbilic is: $x^3 - 3 xy^2 + t_1(x^2 + y^2) + t_2 x + t_3 y$. With determinacy $3$ and codimension $3$ it corresponds to four critical points merging to give a monkey saddle. When $t_2 = t_3 = 0$, it possesses three-fold rotation symmetry ($\rho_{2\pi /3}$) along with $r_y$.

\subsubsection{Hyperbolic umbilic}
The hyperbolic umbilic is similar to the elliptic umbilic except for a sign difference; the unfolding is: $x^3 + 3 xy^2 + t_1 x^2 + t_2 x + t_3 y$. It too has determinacy $3$ and codimension $3$. However it does not possess the three-fold rotational symmetry of the monkey saddle: the only symmetry is $r_y$ when $t_3 = 0$.

\subsubsection{Parabolic umbilic}
The parabolic umbilic has determinacy $4$ and codimension $4$. It's standard universal unfolding is given by: $\pm(x^2 y + y^4 + t_1 y^2 + t_2 x^2 + t_3 y + t_4 x)$. Under $t_4 = 0$, it has $r_x$ as a symmetry.

We have illustrated Thom's theorem in two dimensions. For $n$ dimensions, we simply add a Morse part $\pm y_1^2 \pm \cdots \pm y_{n-2}^2$ to the unfoldings given. Thoms theorem can be extended to $r=6$ for arbitrary $n$ and $r=7$ for $n=2$. This requires us to incorporate all of the catastrophes with codimension $\leqslant r$. This has been done in Tables~\ref{table2} and~\ref{table3}. In addition to these, we consider one other singularity which is allowed by $\pi/4$ rotation symmetry:
\subsubsection{$X_9$}
This is actually a class of mutually inequivalent singularities of the form $x^4 + 2cx^2 y^2 +y^4$. For $|c| \neq 1$ this is a higher order singularity with codimension $8$ and determinacy $4$. For two different parameters $c$ and $c'$, the singularities are equivalent if and only if $c,c' > -1$ and $c' = (3-c)/(1+c)$. The function itself has a four-fold rotational symmetry and we will use a symmertry consistent unfolding: $\epsilon = x^4 + 2cx^2 y^2 +y^4 + t(x^2 + y^2)$.

\section{Taylor expansions and catastrophes}
A family which is smoothly equivalent to the standard universal unfolding of a catastrophe can have a Taylor expansion that does not resemble the unfolding to any order. In fact, it may not be obvious what catastrophe is equivalent to the given family by simply looking at the Taylor expansion truncated to any degree. This complication presents not just for families and catastrophes, but also for functions and standard forms of higher order singularities. Given a function with a higher order critical point at the origin, it is in general very hard to find the local diffeomorphism that transforms it to the standard form. So in practice, it becomes necessary to avoid finding explicit coordinate transformations in order to determine the type of higher order critical point. \\

As seen earlier, \emph{corank, codimension, determinacy, winding} are preserved under a smoothly reversible coordinate transformation and the standard types are described uniquely by these numbers. Thus if we calculate these numbers for a given function, we can unamibguisly identify the type of higher order singularity it is equivalent to. It is fairly easy to compute the corank and winding of a function (see Appendix~\ref{sec:corank} \&~\ref{sec:winding}). The problem therefore reduces to computing the determinacy and codimension of the given function in an efficient way. This is achieved by the method given in Sec.~\ref{sec:algorithm} where we also apply the method to an example. We now explain the reason behind the method.\\

\subsection{Explanation of the method}\label{app:method_explanation}

For any function $f$, the $k^{\mathrm{th}}$ degree Taylor polynomial $j^k f$ moves in the space of all polynomials under a smooth coordinate transformation. For example, let $j^3 f = k_x^3 - k_y^2$. Under $k_x \rightarrow k_x$ and $k_y \rightarrow k_y-k_x^2$, we have $j^3 f \rightarrow k_x^3 - k_y^2 + 2k_x^2 k_y$. Assume that the critical point of the function occurs at the origin. By applying a family of origin preserving smooth coordinate transformations $\phi(t)$ parametrized by a continuous real number $t$, we can generate the orbits $j^k [f(\phi(t))]$ of $j^k f$ in the polynomial space. (We assume that $\phi(0)$ is the identity coordinate transformation and refer to the original Taylor polynomial at the origin $j^k[f(\phi(0))]$ as simply $j^k f$). The orbits are essentially curves $\gamma(t)$ parametrized by $t$ with $\gamma(0) = j^k f$. We can define tangents to these curves at $j^k f$: 
\begin{equation}
\gamma'(0) = \left.\frac{d}{dt} j^k[f(\phi(t))]\right\vert_{t = 0}
\end{equation}
The space of such tangents for all possible origin preserving one parameter families is the tangent space at $j^k f$. An important result in catastrophe theory guarantees the following: if the tangent space at $j^k f$ contains all homogeneous monomials of the form $k_x^j k_y^{k-j}$ for $j = 0,\cdots,k$, then terms of $O(k+1)$ can be removed from $f$ by a smooth coordinate transformation. This ensures that $f$ is $k$-determinate. Conversely, if $f$ is $(k-1)$-determinate, then the tangent space at $j^{k} f$ contains all the homogeneous monomials $k_x^j k_y^{k-j}$ for $j = 0,\cdots,k$. By determining the lowest $k$ for which the tangent space at $j^k f$ contains $k_x^j k_y^{k-j}$ for $j = 0,\cdots,k$ we can narrow down the determinacy to either $k$ or $k-1$. To this end, in the method in Sec.~\ref{sec:algorithm} we generate a set of polynomials $\{p_i \}$ that span the tangent space and try to take linear combinations of these polynomials to generate $k_x^j k_y^{k-j}$ for $j = 0,\cdots,k$.\\

If we also allow smooth coordinate transformations that move the origin, the tangent space gets enlarged. The codimension of $f$ is then codimension of this enlarged tangent space at $j^k f$ in the space of \emph{all} polynomials with zero constant term and degree $\leqslant k$, for any $k$ for which $f$ is $k$ determinate. Thus by finding a $k$ for which $f$ is $k$-determinate and generating a spanning set for the enlarged tangent space, we can find the codimension by finding the dimension of the tangent space and subtracting it from the dimension of the polynomial space. To find the dimension of the tangent space, we just list the spanning vectors as the rows of a matrix in the basis of monomials and compute the rank of this matrix.\\


\section{Isometries of the plane}
\label{app:isometries}
This section is mostly based on the treatment in~\cite{artin2011algebra}. Let $\mathsf{P}$ denote the two-dimensional plane. When an origin is chosen, $\mathsf{P}$ can be identified with $\mathbb{R}^2$ so that any vector $\bm{u}$ can be written in terms of two real components as $(u_1, u_2)$ (here on, we will use $\mathsf{P}$ and $\mathbb{R}^2$ interchangeably). For two vectors $\bm{x} = (x_1, x_2)$ and $\bm{y} = (y_1, y_2)$ let $d(\bm{x}, \bm{y}) = \sqrt{(x_1 - x_2)^2 + (y_1 - y_2)^2}$ denote the euclidean distance between them. The euclidean distance can also be obtained from the dot product: $\bm{x} \cdot \bm{y} = x_1 y_1 + x_2 y_2$. An isometry of the plane is a function $f: \mathsf{P} \rightarrow \mathsf{P}$ which preserves the euclidean distance. i.e., $d(f(\bm{x}), f(\bm{y})) = d(\bm{x}, \bm{y})$. \\ \\
We denote a copy of $\mathbb{R}^2$ as $\mathsf{V}$, the group of all translations (it is a group under addition of vectors). It acts on $\mathsf{P}$ as follows: If $\bm{a} \in \mathsf{V}$, the translation by $\bm{a}$, denoted by $t_{\bm{a}}: \mathsf{P} \rightarrow \mathsf{P}$, simply adds $\bm{a}$ to the vectors in $\mathsf{P}$: $t_{\bm{a}} (\bm{x}) = \bm{x} + \bm{a}$. It is important to note that $\mathsf{V}$ has a fixed origin while for $\mathsf{P}$, we can choose any point as the origin. The following are equivalent definitions of an orthogonal operator $\phi : \mathbb{R}^2 \rightarrow \mathbb{R}^2$~\cite{artin2011algebra} \\
(a) $\phi$ is an isometry that preserves the origin: $\phi(0) = 0$, \\
(b) $\phi$ preserves dot products: $\phi (\bm{x}) \cdot \phi (\bm{y}) = \bm{x} \cdot \bm{y}$, \\
(c) The matrix of $\phi$, say $A$, is such that it's transpose is it's inverse $A^T = A^{-1}$. \\ \\
Any orthogonal operator in two dimensions is either a rotation by some angle $\theta$, which we denote by $\rho_{\theta}$ or a reflection about some line through the origin. In the latter case, it can be uniquely decomposed as the composition of a rotation $\rho_{\theta}$ and reflection about the $x$-axis (which we shall denote by $r$): $\phi = \rho_{\theta} \, r$ (see below). It is clear from above, that orthogonal operators act on $\mathsf{P}$ only after an origin has been chosen. Also, it is easily seen that translations and orthogonal operators are isometries. In fact, it can be shown that any isometry $f$ can be \emph{uniquely} decomposed as the composition of a translation and an orthogonal operator: $f = t_{\bm{a}} \phi$~\cite{artin2011algebra}. Moreover, the set of all isometries of the plane is a group with the composition of functions as it's group law of composition. We denote this by $M$. It contains translations, rotations about points (not just the origin), reflections about lines and glide reflections.

\subsection{Compositions of basic isometries}
Using the matrix form of $\rho_{\theta}$ and $r$, we can easily show that $r \rho_{\theta} = \rho_{-\theta} \, r$. The operator $\rho_{2\theta} \, r$ is actually a reflection about the line through the origin which makes an angle $\theta$ with the $x$-axis. The easiest way to see this is to do a coordinate transformation $U=\rho_{-\theta}$ which rotates the plane by $-\theta$ so that the line in consideration becomes the $x$-axis. Under this, an isometry $\phi$ transforms as $\phi \rightarrow U \, \phi \, U^{-1}$. In particular we have $\rho_{2\theta} \, r \rightarrow \rho_{-\theta} \, \rho_{2\theta} \, r \rho_{\theta} = \rho_{-\theta} \, \rho_{2\theta} \, \rho_{-\theta} \, r = r$. Thus, in the new coordinate system the isometry is the reflection about $x$-axis so that in the old coordinate system it is the reflection about the $\theta$-line through the origin. Next we show that the composition of two reflections is a rotation: $r \rho_{\theta} \, r \rho_{\alpha} = rr \rho_{-\theta} \, \rho_{\alpha} = \rho_{\alpha - \theta}$. \\ \\
Now we investigate the composition of a translation and an orthogonal operator. We first note the following: $\rho_{\theta} \, t_{\bm{a}} = t_{\rho_{\theta} (\bm{a})} \, \rho_{\theta}$. To show that the composition of translation $t_{\bm{a}}$ and rotation $\rho_{\theta}$ is a rotation about some point $\bm{b}$ by $\theta$, we need to show that the equation $t_{\bm{a}} \, \rho_{\theta} = t_{\bm{b}} \, \rho_{\theta} \, t_{-\bm{b}}$ has a unique solution for $\bm{b}$. (The right hand side amounts to shifting $\bm{b}$ to the origin, rotating the plane by $\theta$ and shifting back by $\bm{b}$ which is the same as rotating by $\theta$ \emph{about} $\bm{b}$). Since $t_{\bm{b}} \, \rho_{\theta} \, t_{-\bm{b}} = t_{\bm{b} - \rho_{\theta}(\bm{b})}$, the equation reduces to $\bm{a} = (1-\rho_{\theta}) \, \bm{b}$, which is a pair of linear equations in two unknowns (the two components of $\bm{b}$). This has a unique solution if and only if $\det (1-\rho_{\theta}) \neq 0$. But $\det (1 - \rho_{\theta}) = 2 - 2 \cos \theta$. For $\theta \in [0, 2\pi)$, only $\theta = 0$ satisfies $2 - 2\cos \theta = 0$. Thus, for any $\theta \neq 0$ (mod $2\pi$), we have a unique non-zero $\bm{b}$ corresponding to each $\bm{b} \neq 0$. This verifies the assertion that any composition of rotation and translations is a rotation by the same angle about some point. \\ \\
Before we look at the composition of translations and reflections about lines through the origin (which take the form $\rho_{\theta} \, r$), we rotate the plane to make the reflection axis the $x$-axis. We first consider the composition of a $y$-translation with the $x$-reflection: $t_{a \hat{\bm{j}}} \, r = t_{\frac{a}{2} \hat{\bm{j}}} \, t_{\frac{a}{2} \hat{\bm{j}}} \, r = t_{\frac{a}{2} \hat{\bm{j}}} \, r \, t_{-\frac{a}{2} \hat{\bm{j}}}$. But the latter expression really performs reflection about the line $y = a/2$ (since the operation shifts this line to the $x$-axis, performs $x$-reflection and then shifts backs the line). Thus, the composition of a reflection with translation by a vector perpendicular to the reflection axis amounts to simply shifting the reflection axis by half the translation vector. For a general translation vector, we split the vector into a parallel and perpendicular parts (with respect to the reflection axis). The perpendicular part shifts the reflection axis while the parallel part adds a glide term: $t_{\bm{a}} \, r = t_{\bm{a}_{\parallel}} \, t_{\bm{a}_{\perp}} \, r = t_{\bm{a}_{\parallel}} \, (t_{\bm{a}_{\perp}/2} \, r \, t_{-\bm{a}_{\perp}/2})$. Thus, a generic composition of a translation and a reflection is a glide reflection. As before, the composition of two glide reflections is a rotation.

\subsection{Discrete subgroups of isometries}
A subgroup $G$ of $M$ is discrete if it does not contain arbitrarily small rotations and translations. Mathematically, this means that there is a positive real number $\epsilon$ such that for any translation $t_{\bm{a}} \in G$, we have $\| \bm{a} \| < \epsilon$ and for any rotation about some point, we have the angle of rotation $\theta < \epsilon$. \\ \\
Given the unique decomposition of an isometry as $f = t_{\bm{a}} \phi$, we can define a map $\pi : M \rightarrow O(2)$, $\pi (t_{\bm{a}} \phi) = \phi$. This is a group homomorphism. We now restrict $\pi$ to $G$. The image of $G$ under $\pi$, denoted by $\bar{G}$ is a discrete subgroup of $O(2)$. This is called the point group. An element of point group \emph{need not} be an element of $G$, so that in general $\bar{G}$ is \emph{not} a subgroup of $G$. (This can happen for instance if $G$ contains only glide reflections and no pure reflections. But the point group will contain a pure reflection). Thus, the point group contains information about the orthogonal operations present in $G$, either as pure orthogonal operators or in combination with translations. We can show that the only discrete subgroups of $O(2)$ are the finite subgroups: the cyclic subgroups $C_n$ and the dihedral groups $D_n$. The cycilc subgroup $C_n$ of order $n$ is generated by the rotation $\rho_{\theta}$ with $\theta = 2\pi /n$ while the dihedral group $D_n$, of order $2n$ is generated by $\rho_{\theta}$ and $r'$, where $\theta = 2\pi /n$ and $r'$ is a reflection about a line through the origin. Thus, for any discrete group of isometries $G$, the point group $\bar{G}$ is $C_n$ or $D_n$ for a positive integer $n$. \\ \\
There is one other group assosiated with a dicrete group of isometries: the lattice, denoted by $\mathsf{L}$. It contains all the translations contained in $G$. In contrast to the point group, the lattice is indeed a subgroup of $G$. The lattice could be: \\
(a) the zero group $\{ 0 \}$, \\
(b) the set of integer multiples of a non-zero vector $\bm{a}$: $\mathsf{L} = \{m \bm{a} \, | \, m \in \mathbb{Z} \}$, \\
(c) the set of integer combinations of two independent vectors $\bm{a}$ and $\bm{b}$: $\mathsf{L} = \{ m\bm{a} + n\bm{b} \, | \, m, n \in \mathbb{Z} \}$. \\ \\
There are precisely seventeen groups which have lattices of the form given in (c). They are known as the wallpaper groups and we shall restrict our interest to them. When the case (c) occurs, the point group is restricted to be $C_n$ or $D_n$ with $n = 1,2,3,4$ or $6$. This is known as crystallographic restriction. Another important property that ties up $\mathsf{L}$ and $\bar{G}$ is that $\bar{G}$ preserves $\mathsf{L}$. In other words, for any $\phi \in \bar{G}$, $\phi(\bm{a}) \in \mathsf{L} \,\, \forall \bm{a} \in \mathsf{L}$ or $\phi(\mathsf{L}) = \mathsf{L}$.

\subsection{Wallpaper groups}
There are seventeen wallpaper groups in two-dimensions~\cite{artin2011algebra}. Each of these 17 groups has one or more associated lattices drawn from the five basic types: square, rectangle, hexagonal, rhombic (also known as centered rectangle) and oblique. The lattice and point group do not, in general determine the wallpaper group uniquely although, as we shall show, for our purposes they are sufficient. Typically, one encounters wallpaper groups as the symmetry groups of two dimensional crystals. More precisely, if $V(x)$ is the potential experienced by the electrons in the non-interacting picture and $G$ is the wallpaper group associated with the underlying crystal, then for any symmetry $S \in G$, $V(S\bm{x}) = V(\bm{x})$. This has important consequences for the energy spectrum, one of which is Bloch's theorem.

\subsection{Symmetries and Schr{\"o}dinger equation}
Let $G$ be one of the wallpaper groups and let $S \in G$. The unique decomposition of $S$ is $t_{\bm{a}} \phi$ so that $\tilde{\bm{r}} = S\bm{r} = \phi(\bm{r}) + \bm{a}$. We denote the matrix of $\phi$ by $\phi$ as well. Now,
\begin{multline}
\tilde{r}_{i} = \phi_{ij} \, r_j + a_i \Rightarrow \frac{\partial f}{\partial r_j} = \frac{\partial f}{\partial \tilde{r}_i} \, \frac{\partial \tilde{r}_i}{\partial r_j} = \phi_{ij} \, \frac{\partial f}{\partial \tilde{r}_i} \\ \Rightarrow \nabla_{\bm{r}}^2 f = \sum_j \frac{\partial^2 f}{\partial r_j^2} = \sum_j \phi_{ij} \frac{\partial}{\partial r_j} \bigg( \frac{\partial f}{\partial \tilde{r}_i} \bigg) \\
= \sum_j \phi_{ij} \frac{\partial^2 f}{\partial \tilde{r}_k \partial \tilde{r}_i} \phi_{kj} = \delta_{ik} \frac{\partial^2 f}{\partial \tilde{r}_k \partial \tilde{r}_i} = \sum_k \frac{\partial^2 f}{\partial \tilde{r}_k^2} = \nabla_{\tilde{\bm{r}}}^2 f
\end{multline}
Thus, the gradient is invariant under an isometry. Now consider the time independent Schr{\"o}dinger equation, incorporating Bloch's theorem:
\begin{equation}
\bigg( \frac{-\hbar^2}{2m} \nabla_{\bm{r}}^2 + V(\bm{r}) \bigg) \psi_{n, \bm{k}}(\bm{r}) = \epsilon_n (\bm{k}) \psi_{n, \bm{k}}(\bm{r})
\end{equation}
Now this holds good under a relabeling $\bm{r} \rightarrow \tilde{\bm{r}}$ so that
\begin{multline}
\bigg( \frac{-\hbar^2}{2m} \nabla_{\tilde{\bm{r}}}^2 + V(\tilde{\bm{r}}) \bigg) \psi_{n, \bm{k}}(\tilde{\bm{r}}) = \epsilon_n (\bm{k}) \psi_{n, \bm{k}}(\tilde{\bm{r}}) \\ \Rightarrow \bigg( \frac{-\hbar^2}{2m} \nabla_{\bm{r}}^2 + V(\bm{r}) \bigg) \psi_{n, \bm{k}}(\tilde{\bm{r}}) = \epsilon_n (\bm{k}) \psi_{n, \bm{k}}(\tilde{\bm{r}})
\end{multline}
Since $\nabla_{\tilde{\bm{r}}}^2 = \nabla_{\bm{r}}^2$ and $V(\tilde{\bm{r}}) = V(\bm{r})$. We define $\tilde{\psi}_{n, \bm{k}}(\bm{r}) = \psi_{n, \bm{k}} (\tilde{\bm{r}}) = \psi_{n, \bm{k}} (\phi (\bm{r}) + \bm{a})$. This gives:
\begin{equation}
\bigg( \frac{-\hbar^2}{2m} \nabla_{\bm{r}}^2 + V(\bm{r}) \bigg) \tilde{\psi}_{n, \bm{k}}(\bm{r}) = \epsilon_n (\bm{k}) \tilde{\psi}_{n, \bm{k}}(\bm{r})
\end{equation}
So that $\tilde{\psi}_{n, \bm{k}}(\bm{r})$ also has eigenvalue $\epsilon_n(\bm{k})$. To find its crystal momentum we evaluate $\tilde{\psi}_{n, \bm{k}}(\bm{r} + \bm{R})$:
\begin{multline}
\tilde{\psi}_{n, \bm{k}}(\bm{r} + \bm{R}) = \psi_{n, \bm{k}}(\phi(\bm{r} + \bm{R}) + \bm{a}) = \psi_{n, \bm{k}}(\phi(\bm{r}) + \bm{a} + \phi(\bm{R}))\\ = e^{i \bm{k} \cdot \phi(\bm{R})} = e^{i \phi^{-1} (\bm{k}) \cdot \bm{R}} \tilde{\psi}_{n, \bm{k}}(\bm{r})
\end{multline}
Thus the crystal momentum of $\tilde{\psi}_{n, \bm{k}}(\bm{r})$ is $\phi^{-1}(\bm{k})$ or $\tilde{\psi}_{n, \bm{k}}(\bm{r}) = \psi_{n, \phi^{-1}(\bm{k})} (\bm{r})$. This gives $\epsilon(\phi^{-1} (\bm{k})) = \epsilon (\bm{k})$. This means that the point group is a symmetry group for the dispersion $\epsilon_n(\bm{k})$, alongside the reciprocal lattice translations.

%
\end{document}